\documentclass[11pt,a4paper]{article}
\pdfoutput=1
\usepackage{jheppub}
\usepackage{color,epsfig} 
\usepackage{ifpdf}
\usepackage{amsmath} 
\usepackage{amssymb} 
\usepackage{bm}
\usepackage[english]{babel}
\usepackage{slashed}
\usepackage{multirow}
\usepackage{pbox}
\usepackage{subfigure}

\title{Asymmetric Leptoquark Pair Production at LHC}

\author[a,b]{Ilja Dor\v sner,} 
\author[c]{Ajla Lejli\' c,}
\author[d]{Shaikh Saad} 

\affiliation[a]{University of Split, Faculty of Electrical Engineering, Mechanical Engineering and Naval Architecture in Split, Ru\dj era Bo\v skovi\' ca 32, HR-21000 Split, Croatia}
\affiliation[b]{J.\ Stefan Institute, Jamova 39, P.\ O.\ Box 3000, SI-1001
  Ljubljana, Slovenia}
\affiliation[c]{Department of Physics, Faculty of Science,
  University of Zagreb, Bijeni\v cka cesta 32, HR-10000 Zagreb, Croatia}
\affiliation[d]{Department of Physics, University of Basel, Klingelbergstrasse\ 82, CH-4056 Basel, Switzerland}  
  
\emailAdd{dorsner@fesb.hr}
\emailAdd{ajla@me.com} 
\emailAdd{shaikh.saad@unibas.ch}

\abstract{
We investigate asymmetric leptoquark pair production mechanism at the Large Hadron Collider to advocate its potential relevance to establish reliable constraints on the leptoquark parameter space and its ability to aid in correct identification of these attractive sources of new physics. The main feature of asymmetric pair production that genuinely distinguishes it from the usual leptoquark pair production is given by the fact that the two leptoquarks that are produced in proton-proton collisions through a $t$-channel lepton exchange are not charge conjugates of each other. Hence the proposed name of asymmetric leptoquark pair production for this type of process. We spell out prerequisite conditions for the asymmetric leptoquark pair production mechanism to be operational and enumerate all possible combinations of leptoquark multiplets that can potentially generate it. We finally reinterpret existing leptoquark pair production search results within several simple scalar leptoquark extensions of the Standard Model, assuming that the leptoquarks exclusively couple to either electrons or muons and the first generation quarks, to demonstrate proper inclusion of asymmetric pair production. We consequently present accurate parameter space constraints for the $S_1$, $S_3$, $R_2$, $S_1$+$S_3$, and $S_1$+$R_2$ leptoquark scenarios.}

\begin{document}

\maketitle
\section{Introduction}
\label{sec:introduction}

Leptoquark pair production is the only available process to efficiently look for these hypothetical particles at hadron colliders when the coupling strength between the relevant quark-lepton pairs and a leptoquark is small. It is thus clear that the search for leptoquarks via pair production is always going to be an integral part of the Large Hadron Collider (LHC) experimental agenda in years, if not decades, to come. (For a sample of the leptoquark pair production search results, see Refs.~\cite{ATLAS:2020dsk,ATLAS:2020xov,CMS:2020wzx,ATLAS:2021oiz,CMS:2022zks}.) The leptoquark pair production cross sections applicable to LHC are accordingly available at the next-to-leading order~\cite{Kramer:1997hh,Kramer:2004df,Mandal:2015lca,Dorsner:2018ynv} as well as the next-to-next-to-leading order~\cite{Beenakker:2016lwe,Beenakker:1997ut,Beenakker:2010nq,Beenakker:2016gmf} in strong coupling constant, and, more recently, at the next-to-leading order in both the strong coupling constant and the leptoquark Yukawa coupling(s)~\cite{Borschensky:2020hot}.

As the strength of interaction between the quark-lepton pairs and a leptoquark is gradually increased, the collider searches for signals from several other processes start to be relevant in constraining the leptoquark parameter space. These processes, at the LHC, are a single leptoquark production~\cite{Alves:2002tj,Dorsner:2014axa,Hammett:2015sea,Mandal:2015vfa,Dorsner:2018ynv,Schmaltz:2018nls}, a non-resonant production of the Drell-Yan type~\cite{Faroughy:2016osc,Raj:2016aky,Greljo:2017vvb,Bansal:2018eha,Schmaltz:2018nls,Fuentes-Martin:2020lea,Allwicher:2022gkm}, and a resonant leptoquark production~\cite{Ohnemus:1994xf,Eboli:1997fb,Buonocore:2020erb,Greljo:2020tgv,Buonocore:2022msy}. 

Note, however, that even the leptoquark pair production exhibits dependence on the Yukawa coupling strength~\cite{Dorsner:2014axa}. This is especially true in the case of a novel mechanism of leptoquark pair production that has been recently introduced in Ref.~\cite{Dorsner:2021chv}. The main feature of this novel mechanism that distinguishes it from the usual leptoquark pair production at the LHC is the fact that the two leptoquarks that are produced in proton-proton collisions through a $t$-channel exchange of a lepton do not comprise a charge conjugate pair. This is a primary reason why we refer to it as an asymmetric leptoquark pair production in this study. The novel production mechanism, though, can yield the same final state as the conventional pair production. In fact, the final state kinematics should be exactly the same if the leptoquarks in question are degenerate in mass. 
This work aims to address the correct interpretation of existing and future experimental search results for those final states that are due to the leptoquark pair production processes and subsequent leptoquark decays if one appropriately incorporates the aforementioned asymmetric mechanism contributions. It dovetails the initial analysis of Ref.~\cite{Dorsner:2021chv} and extends the scope of the phenomenological discussion of asymmetric pair production presented therein. It also nicely complements recent work on the inclusion of the asymmetric pair production mechanism in the next-to-leading order in QCD cross section determinations for the leptoquark pair production~\cite{Borschensky:2022xsa}. We stress that the asymmetric contributions to the leptoquark pair production have not been included in any of publicly available experimental search analyses thus far.

We will, for definiteness, focus our attention solely on the scalar leptoquark extensions of the Standard Model (SM). We accordingly present in Table~\ref{tab:list} a list of pertinent scalar leptoquarks and associated transformation properties under the SM gauge group $SU(3) \times SU(2) \times U(1)$. Since the chirality of the leptons that the scalar leptoquark couples to is very important for our discussion, we indicate relevant chiralities of both quarks and leptons using $R$ and $L$ for right- and left-chiral fields, respectively, in the third column of Table~\ref{tab:list}. Our convention is such that the first (second) letter, in that column, denotes chirality of quarks (leptons). For example, the fact that $S_1$ leptoquark can directly couple to the $SU(2)$ doublets of quarks and leptons or/and the $SU(2)$ singlets of quarks and leptons is indicated by simultaneous presence of $LL$ and $RR$ designations in the third column of Table~\ref{tab:list}.
\begin{table}[tbp]
\centering
\begin{tabular}{|c|c|c|c|}
\hline
$(SU(3),SU(2),U(1))$ &  LQ SYMBOL & CHIRALITY TYPE (LQ-$q$-$l$) & $F$ \\
\hline 
\hline
$(\overline{\mathbf{3}},\mathbf{3},1/3)$ & $S_3$ & $LL$ & $-2$ \\
$(\mathbf{3},\mathbf{2},7/6)$ & $R_2$ & $RL$, $LR$ & $0$ \\
$(\mathbf{3},\mathbf{2},1/6)$ & $\tilde{R}_2$ & $RL$ & $0$ \\
$(\overline{\mathbf{3}},\mathbf{1},4/3)$ & $\tilde{S}_1$ & $RR$ & $-2$ \\
$(\overline{\mathbf{3}},\mathbf{1},1/3)$ & $S_1$ & $LL$, $RR$ & $-2$ \\
\hline 
\end{tabular}
\caption{Scalar leptoquark multiplets, chiralities of the leptoquark interactions with the SM quark-lepton pairs, and associated leptoquark fermion numbers.}
\label{tab:list} 
\end{table}
We also specify fermion number $F$ of scalar leptoquark multiplets in Table~\ref{tab:list}, where $F$ is defined as the sum of the lepton number and three times the baryon number of leptons and quarks that a given leptoquark couples to. Leptoquarks with $F=-2$ exclusively couple/decay to quarks and leptons whereas $F=0$ leptoquarks couple/decay to quark-antilepton or antiquark-lepton pairs.

Since our hyper-charge normalization is $Q=I_3+Y$, where $Q$ corresponds to electric charge in units of the positron charge, $I_3$ stands for the diagonal generator of $SU(2)$, and $Y$ represents $U(1)$ hyper-charge operator, the  electric charge eigenvalues of scalar leptoquarks in Table~\ref{tab:list} are $S_3^{+4/3}$, $S_3^{+1/3}$, $S_3^{-2/3}$, $R_2^{+5/3}$, $R_2^{+2/3}$, $\tilde{R}_2^{+2/3}$, $\tilde{R}_2^{-1/3}$, $\tilde{S}_1^{+4/3}$, and $S_1^{+1/3}$. We will always denote leptoquarks using this notation and furthermore write, for simplicity, that $(\mathrm{LQ}^{+Q})^*=\mathrm{LQ}^{-Q}$ and $(\mathrm{LQ}^{-Q})^*=\mathrm{LQ}^{+Q}$. Note that leptoquarks of the same electric charge can, in principle, mix with each other upon the breaking of the SM symmetry down to $SU(3) \times U(1)_\mathrm{em}$ even if they have different fermion numbers. This type of mixing can lead to interesting physical phenomena that are somewhat orthogonal to our study. This is the main reason why we neglect all such possible mixings terms. 

\begin{figure}[th!]\centering
\includegraphics[width=0.9\textwidth]{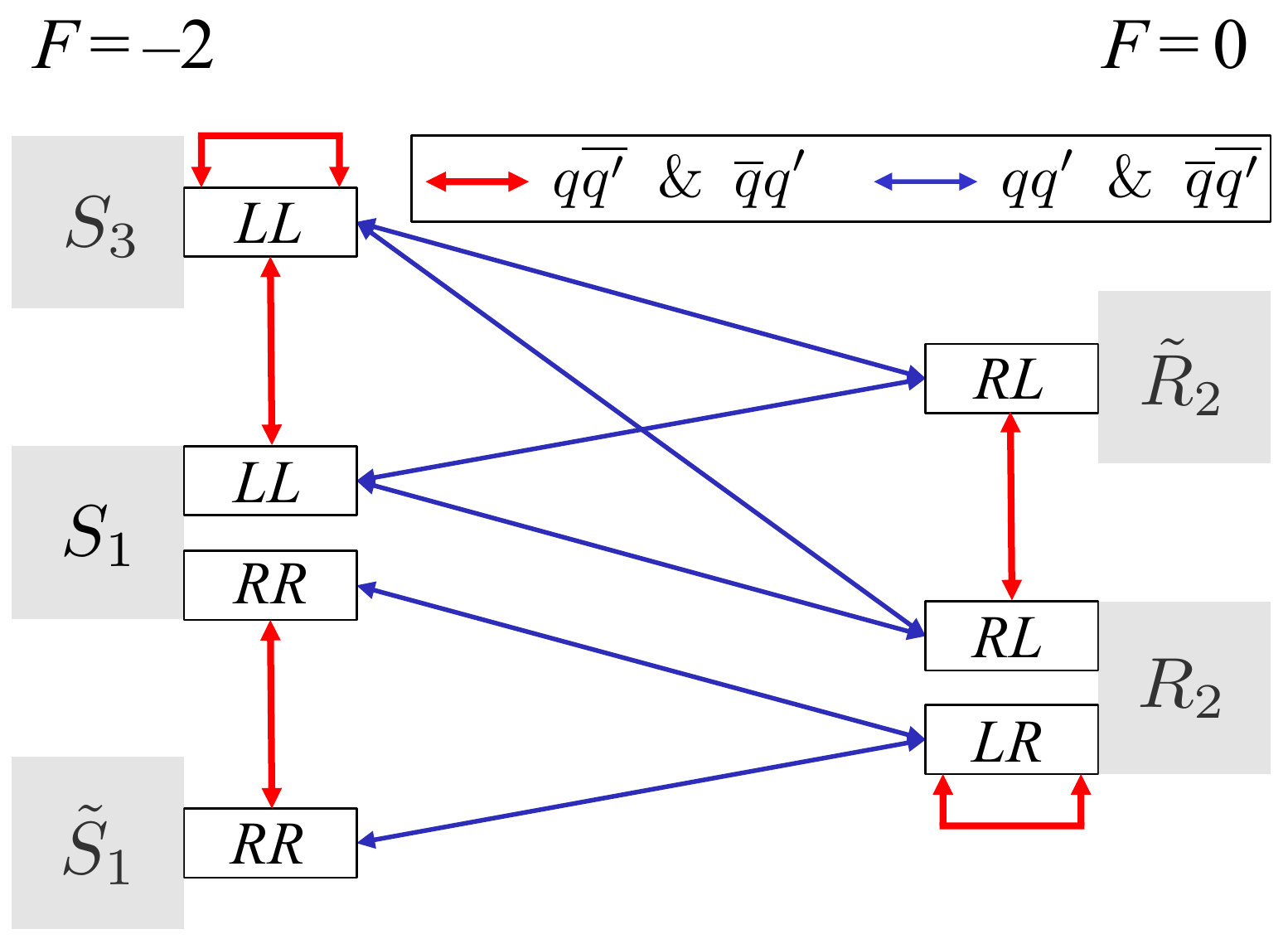}
\caption{Schematic classification of potential sources of asymmetric leptoquark pair production. See text for details.}
\label{fig:LQ_COMBINATIONS}
\end{figure}

There are several prerequisite conditions for the asymmetric leptoquark pair production mechanism under consideration to be operational~\cite{Dorsner:2021chv}. First, it requires non-negligible Yukawa coupling(s) between leptoquarks and the SM quarks and leptons. Second, this mechanism is relevant whenever there exist at least two leptoquark states originating from the same or two different leptoquark multiplets that couple to a lepton of the same chirality and flavor. This, then, leads to a simple schematic representation shown in Fig.~\ref{fig:LQ_COMBINATIONS} of all possible minimal leptoquark combinations that can potentially generate asymmetric pair production at hadron colliders and, consequentially, LHC.

The double-headed arrows in Fig.~\ref{fig:LQ_COMBINATIONS} connect those leptoquark multiplets that can simultaneously couple to a lepton of the same flavor and chirality. For example, $R_2$ can couple to the SM leptons of both chiralities~\cite{Dorsner:2016wpm} as indicated with the $RL$ and $LR$ designations in Fig.~\ref{fig:LQ_COMBINATIONS} and Table~\ref{tab:list}. Again, it is the second letter that denotes the lepton chirality. If $R_2$ couples to the left-chiral leptons, it can, in principle, participate in the asymmetric pair production with all those multiplets that can also couple to the left-chiral leptons such as $S_3$, $S_1$, and $\tilde{R}_2$. If $R_2$ couples to the right-chiral leptons, it can potentially contribute to asymmetric pair production on its own, as indicated in Fig.~\ref{fig:LQ_COMBINATIONS}, and/or in conjunction with $S_1$ and $\tilde{S}_1$. 

The double-headed arrows in Fig.~\ref{fig:LQ_COMBINATIONS} are color-coded either blue or red to distinguish between two different initial state configurations behind the relevant asymmetric pair production processes even though the leptoquark pairs in question are always generated in proton-proton collisions via a $t$-channel lepton exchange. If the two leptoquarks $\mathrm{LQ}_1$ and $\mathrm{LQ}_2$ have different fermion numbers, i.e., $\Delta F=|F(\mathrm{LQ}_1)|-|F(\mathrm{LQ}_2)|=\pm2$, the initial states are of the $q q^{\prime}$ and $\overline{q} \overline{q^{\prime}}$ nature, where $q$ and $q^{\prime}$ denote the quark fields and can, in principle, be equal to $u$, $d$, $s$, $c$, and $b$. These scenarios are indicated with blue double-headed arrows in Fig.~\ref{fig:LQ_COMBINATIONS}. If, on the other hand, leptoquarks have the same fermion number, i.e., $\Delta F=0$, the initial states are of the $q \overline{q^{\prime}}$ and $\overline{q} q^{\prime}$ nature, where, again, $q,q^{\prime}=u,d, s, c, b$. The $\Delta F=0$ scenarios are depicted with red double-headed arrows in Fig.~\ref{fig:LQ_COMBINATIONS}. 
In view of all these requirements, we note that it is entirely possible to have a new physics scenario with only one scalar leptoquark multiplet and only one non-zero Yukawa coupling and still be able to asymmetrically produce leptoquark pairs at the LHC~\cite{Dorsner:2021chv}. There are two such scenarios, as indicated in Fig.~\ref{fig:LQ_COMBINATIONS}. One is generated if a single non-zero Yukawa coupling exists between $R_2$ and any right-chiral charged lepton. The other one requires presence of a single non-zero Yukawa coupling for $S_3$. 

\begin{figure}[th!]\centering
\includegraphics[width=1\textwidth]{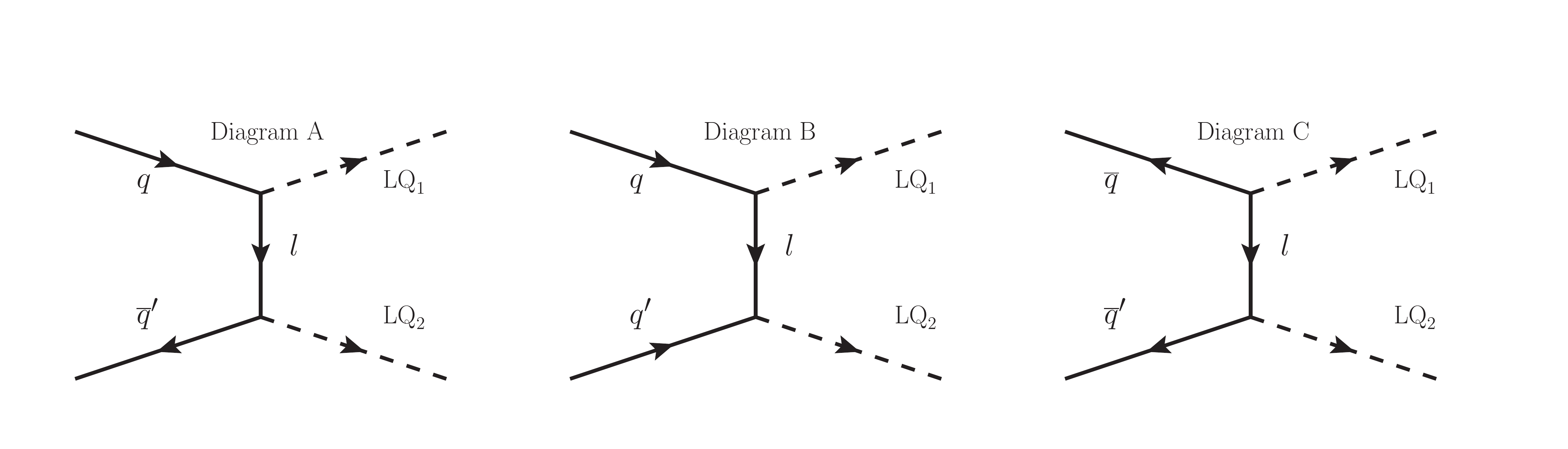}
\includegraphics[width=1\textwidth]{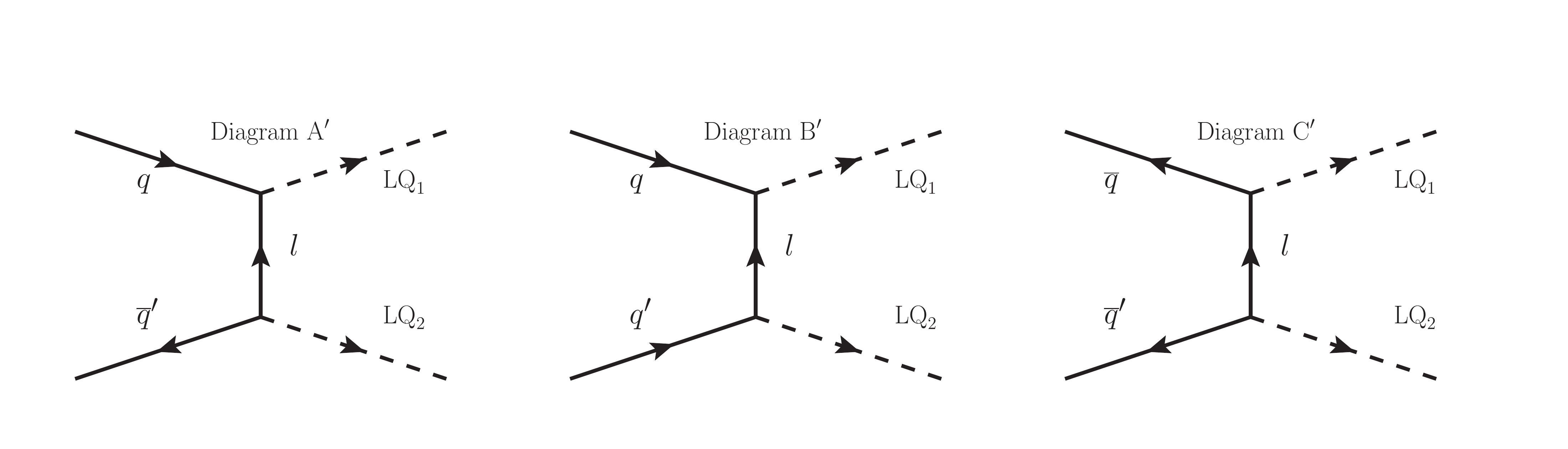}
\caption{Types of diagrams for asymmetric production. $q$, $q^\prime$, $l$, $\mathrm{LQ}_1$, and $\mathrm{LQ}_2$ for $\Delta F=|F(\mathrm{LQ}_1)-|F(\mathrm{LQ}_2)|=\pm 2$ and $\Delta F=|F(\mathrm{LQ}_1)|-|F(\mathrm{LQ}_2)|=0$ are specified in Tables~\ref{tab:101} and \ref{tab:102}, respectively. Here $l$ refers to a charged lepton or a neutrino.}\label{diag-type}
\end{figure}

It is possible to succinctly depict all the relevant diagrams that result in asymmetric pair production at hadron colliders. There are, all in all, six such $t$-channel diagrams that can potentially generate asymmetric pair production. We present these diagrams in Fig.~\ref{diag-type} and then summarize in  
Tables~\ref{tab:101} and \ref{tab:102} associated scenarios that require presence of, at most, two scalar leptoquark multiplets when $\Delta F=0$ and $\Delta F=\pm 2$, respectively. 

\begin{table}[th!]
\begin{center}
\begin{tabular}{|c|c|c|c|c|c|c|}\hline
Diagram Type & $q$ & $\overline{q^{\prime}}$ & $l$ & LQ$_1$ & LQ$_2$& LQ scenario  \\\hline\hline

A&$u_L$&$\overline d_L$&$\ell_R$&$R_2^{+5/3}$&$R_2^{+2/3}$ & \multirow{2}{1cm}{~~$R_2$}\\ \cline{1-6}
A&$d_L$&$\overline u_L$&$\ell_R$&$R_2^{-2/3}$&$R_2^{-5/3}$& \\ \hline\hline

A$^{\prime}$&$d_L$&$\overline u_L$&$\nu_L$&$S_3^{-1/3}$&$S_3^{-2/3}$ & \multirow{4}{1cm}{~~$S_3$}\\ \cline{1-6}
A$^{\prime}$&$u_L$&$\overline d_L$&$\nu_L$&$S_3^{+2/3}$&$S_3^{+1/3}$& \\ \cline{1-6} 
A$^{\prime}$&$d_L$&$\overline u_L$&$\ell_L$&$S_3^{-4/3}$&$S_3^{+1/3}$& \\ \cline{1-6}
A$^{\prime}$&$u_L$&$\overline d_L$&$\ell_L$&$S_3^{-1/3}$&$S_3^{+4/3}$& \\ \hline\hline

A&$d_R$&$\overline u_R$&$\ell_L$&$\widetilde R_2^{+2/3}$&$R_2^{-5/3}$ & \multirow{4}{2cm}{~~$\widetilde R_2$+$R_2$}\\ \cline{1-6}
A&$u_R$&$\overline d_R$&$\ell_L$&$R_2^{+5/3}$&$\widetilde R_2^{-2/3}$& \\ \cline{1-6}
A&$d_R$&$\overline u_R$&$\nu_L$&$\widetilde R_2^{-1/3}$&$R_2^{-2/3}$& \\ \cline{1-6}
A&$u_R$&$\overline d_R$&$\nu_L$&$R_2^{+2/3}$&$\widetilde R_2^{+1/3}$& \\ \hline\hline

A$^{\prime}$&$d_R$&$\overline u_R$&$\ell_R$&$\widetilde S_1^{-4/3}$&$S_1^{+1/3}$ & \multirow{2}{2cm}{~~$\widetilde S_1$+$S_1$}\\ \cline{1-6}
A$^{\prime}$&$u_R$&$\overline d_R$&$\ell_R$&$S_1^{-1/3}$&$\widetilde S_1^{+4/3}$& \\ \hline\hline

A$^{\prime}$&$d_L$&$\overline u_L$&$\nu_L$&$S_1^{-1/3}$&$S_3^{-2/3}$ & \multirow{4}{2cm}{~~$S_1$+$S_3$}\\ \cline{1-6}
A$^{\prime}$&$u_L$&$\overline d_L$&$\nu_L$&$S_3^{+2/3}$&$S_1^{+1/3}$& \\ \cline{1-6}
A$^{\prime}$&$u_L$&$\overline d_L$&$\ell_L$&$S_1^{-1/3}$&$S_3^{+4/3}$& \\ \cline{1-6}
A$^{\prime}$&$d_L$&$\overline u_L$&$\ell_L$&$S_3^{-4/3}$&$S_1^{+1/3}$& \\ \hline

\end{tabular}
\end{center}
\caption{Asymmetric production with $q \overline{q^{\prime}}$ and $\overline{q} q^{\prime}$ initial states. See Fig.~\ref{diag-type} for the diagram type.}
\label{tab:101}
\end{table}

\begin{table}[th!]
\begin{center}
\begin{tabular}{|c|c|c|c|c|c|c|}\hline
Diagram Type & $q/\overline q$ & $q^{\prime}/\overline{q^{\prime}}$ & $l$ & LQ$_1$ & LQ$_2$ &LQ scenario \\\hline\hline

B$^{\prime}$&$d_R$&$u_L$&$\ell_R$&$\widetilde S_1^{-4/3}$&$R_2^{+5/3}$ & \multirow{4}{2cm}{~~$\widetilde S_1$+$R_2$}\\ \cline{1-6}
B$^{\prime}$&$d_R$&$\overline d_L$&$\ell_R$&$\widetilde S_1^{-4/3}$&$R_2^{+2/3}$& \\ \cline{1-6}
C$^{\prime}$&$\overline u_L$&$\overline d_R$&$\ell_R$&$R_2^{-5/3}$&$\widetilde  S_1^{+4/3}$& \\ \cline{1-6}
C$^{\prime}$&$\overline d_L$&$\overline d_R$&$\ell_R$&$R_2^{-2/3}$&$\widetilde  S_1^{+4/3}$& \\ \hline \hline

B&$u_R$&$d_L$&$\nu_L$&$R_2^{+2/3}$&$S_1^{-1/3}$ & \multirow{8}{2cm}{~~$S_1$+$R_2$}\\ \cline{1-6}
B&$u_R$&$u_L$&$\ell_L$&$R_2^{+5/3}$&$S_1^{-1/3}$ & \\ \cline{1-6}
B&$u_L$&$u_R$&$\ell_R$&$R_2^{+5/3}$&$S_1^{-1/3}$ & \\ \cline{1-6}
B&$d_L$&$u_R$&$\ell_R$&$R_2^{+2/3}$&$S_1^{-1/3}$ & \\ \cline{1-6}
C&$\overline d_L$&$\overline u_R$&$\nu_L$&$S_1^{+1/3}$&$R_2^{-2/3}$ & \\ \cline{1-6}
C&$\overline u_L$&$\overline u_R$&$\ell_L$&$S_1^{+1/3}$&$R_2^{-5/3}$ & \\ \cline{1-6}
C&$\overline u_R$&$\overline u_L$&$\ell_R$&$S_1^{+1/3}$&$R_2^{-5/3}$ & \\ \cline{1-6}
C&$\overline u_R$&$\overline d_L$&$\ell_R$&$S_1^{+1/3}$&$R_2^{-2/3}$& \\ \hline\hline

B&$d_R$&$u_L$&$\ell_L$&$\widetilde R_2^{+2/3}$&$S_1^{-1/3}$ & \multirow{2}{2cm}{~~$S_1$+$\widetilde R_2$}\\ \cline{1-6}
C&$\overline u_L$&$\overline d_R$&$\ell_L$&$S_1^{+1/3}$&$\widetilde R_2^{-2/3}$& \\ \hline\hline

B&$u_R$&$d_L$&$\ell_L$&$R_2^{+5/3}$&$S_3^{-4/3}$ & \multirow{6}{2cm}{~~$S_3$+$R_2$}\\ \cline{1-6}
B&$u_R$&$u_L$&$\ell_L$&$R_2^{+5/3}$&$S_3^{-1/3}$ & \\ \cline{1-6}
B&$u_R$&$d_L$&$\nu_L$&$R_2^{+2/3}$&$S_3^{-1/3}$ & \\ \cline{1-6}
C&$\overline d_L$&$\overline u_R$&$\ell_L$&$S_3^{+4/3}$&$R_2^{-5/3}$ & \\ \cline{1-6}
C&$\overline u_L$&$\overline u_R$&$\ell_L$&$S_3^{+1/3}$&$R_2^{-5/3}$ & \\ \cline{1-6}
C&$\overline d_L$&$\overline u_R$&$\nu_L$&$S_3^{+1/3}$&$R_2^{-2/3}$& \\ \hline\hline

B&$d_R$&$d_L$&$\ell_L$&$\widetilde R_2^{+2/3}$&$S_3^{-4/3}$ & \multirow{6}{2cm}{~~$S_3$+$\widetilde R_2$}\\ \cline{1-6}
B&$d_R$&$u_L$&$\ell_L$&$\widetilde R_2^{+2/3}$&$S_3^{-1/3}$ & \\ \cline{1-6}
B&$d_R$&$u_L$&$\nu_L$&$\widetilde R_2^{-1/3}$&$S_3^{+2/3}$ & \\ \cline{1-6}
C&$\overline d_L$&$\overline d_R$&$\ell_L$&$S_3^{+4/3}$&$\widetilde R_2^{-2/3}$ & \\ \cline{1-6}
C&$\overline u_L$&$\overline d_R$&$\ell_L$&$S_3^{+1/3}$&$\widetilde R_2^{-2/3}$ & \\ \cline{1-6}
C&$\overline u_L$&$\overline d_R$&$\nu_L$&$S_3^{-2/3}$&$\widetilde R_2^{+1/3}$& \\ \hline

\end{tabular}
\end{center}
\caption{Asymmetric production with $q q^{\prime}$ and $\overline{q} \overline{q^{\prime}}$ initial states. See Fig.~\ref{diag-type} for the diagram type.}
\label{tab:102}
\end{table}

The schematics in Fig.~\ref{fig:LQ_COMBINATIONS}, diagrams of Fig.~\ref{diag-type}, and Tables~\ref{tab:101} and \ref{tab:102} give a complete classification of the asymmetric pair production processes at hadron colliders. With this exhaustive classification completed we turn our attention towards more quantitative discussion of aforementioned mechanism.

The rest of the manuscript is organised as follows. In Sec.~\ref{sec:MAIN} we address subtleties associated with both the asymmetric and conventional leptoquark pair productions and present several specific instances of inclusion of asymmetric pair production into the usual search strategy for leptoquarks, assuming that the leptoquarks in question exclusively couple to either electrons or muons and the first generation quarks. We consequently present accurate parameter space constraints for the $S_1$, $S_3$, $R_2$, $S_1$+$S_3$, and $S_1$+$R_2$ leptoquark scenarios, where, for the electron coupling case, we generate in Sec.~\ref{sec:APV} the latest limits from the atomic parity violation (APV) searches. We briefly conclude in Sec.~\ref{sec:CONCLUSIONS}.

\section{Asymmetric pair production}
\label{sec:MAIN}

Asymmetric pair production mechanism we want to investigate produces two leptoquarks $\mathrm{LQ}_1$ and $\mathrm{LQ}_2$ that are not charge conjugates of each other through one or more of the $t$-channel diagrams of Fig.~\ref{diag-type}. There is thus no interference between the asymmetric and conventional leptoquark pair productions at the amplitude level even though the final state signatures of both processes, upon the $\mathrm{LQ}_1$ and $\mathrm{LQ}_2$ subsequent decays, can be exactly the same. We can accordingly focus our attention solely on the asymmetric pair production cross sections that can be simply added, if and when appropriate, to the conventional pair production cross sections. We work, for simplicity, at the leading order in QCD and denote the cross sections of interest with 
\begin{equation}
\label{eq:crosssection_a}
\sigma^\mathrm{pair}_{q_1 q_2}(y_{q_1},y_{q_2}, m_{\mathrm{LQ}_1},m_{\mathrm{LQ}_2})=a_{q_1 q_2}( m_{\mathrm{LQ}_1},m_{\mathrm{LQ}_2})|y_{q_1} y_{q_2}|^2,
\end{equation}
where $q_1,q_2=u,\overline{u},d,\overline{d},s,\overline{s},c,\overline{c},b,\overline{b}$. Here, leptoquark $\mathrm{LQ}_i$ of mass $m_{\mathrm{LQ}_i}$ couples to a quark $q_i$ and a lepton $l$ of a given chirality and flavor with strength $y_{q_i}$, where $i=1,2$. 

Note that the cross sections of Eq.~\eqref{eq:crosssection_a} do not depend on whether $\mathrm{LQ}_1$ couples to a quark $q_1$ while $\mathrm{LQ}_2$ couples to a quark $q_2$ or vice versa. This is only relevant for subsequent leptoquark decays. The cross sections of Eq.~\eqref{eq:crosssection_a} also do not depend on the type of lepton that leptoquarks $\mathrm{LQ}_1$ and $\mathrm{LQ}_2$ simultaneously couple to. They are proportional to a square of the product $|y_{q_1}y_{q_2}|$ and can thus be trivially rescaled as a function of Yukawa couplings once they are determined for one particular value of $|y_{q_1}y_{q_2}|$ product. 

We will make an assumption that $\mathrm{LQ}_1$ and $\mathrm{LQ}_2$ are mass-degenerate, i.e., $m_{\mathrm{LQ}_1}=m_{\mathrm{LQ}_2}\equiv m_{\mathrm{LQ}}$, and furthermore take all Yukawa couplings to be real. These two assumptions allow us to introduce cross section $\sigma^\mathrm{pair}_{q_1 q_2}(y_{q_1},y_{q_2}, m_{\mathrm{LQ}})$ that is symmetric in flavor, i.e., $\sigma^\mathrm{pair}_{q_1 q_2} \equiv \sigma^\mathrm{pair}_{q_2 q_1}$, where
\begin{equation}
\label{eq:crosssection_b}
\sigma^\mathrm{pair}_{q_1 q_2}(y_{q_1},y_{q_2}, m_{\mathrm{LQ}})=\sigma^\mathrm{pair}_{q_1 q_2}(y_{q_1},y_{q_2},m_{\mathrm{LQ}_1}=m_{\mathrm{LQ}_2}\equiv m_{\mathrm{LQ}},m_{\mathrm{LQ}_2}).   
\end{equation}
Note that the cross sections of Eq.~\eqref{eq:crosssection_b} allow us to extract limits on the leptoquark parameter space from existing experimental searches in a straightforward fashion since the current analyses rely on an explicit assumption of mass degeneracy for hypothetical leptoquark pairs being produced.

There are fifteen cross sections $\sigma^\mathrm{pair}_{q_1 q_2}(y_{q_1},y_{q_2}, m_{\mathrm{LQ}})$ of interest, at the LHC, when the initial states are quark-quark pairs and twenty five when the initial states are quark-antiquark pairs. We are not interested in the cross sections that are antiquark-antiquark initiated as these are highly suppressed at the LHC although we include them for completeness in the numerical simulation once we reinterpret current leptoquark search analyses results. 

\begin{figure}[th!]\centering
\includegraphics[width=0.95\textwidth]{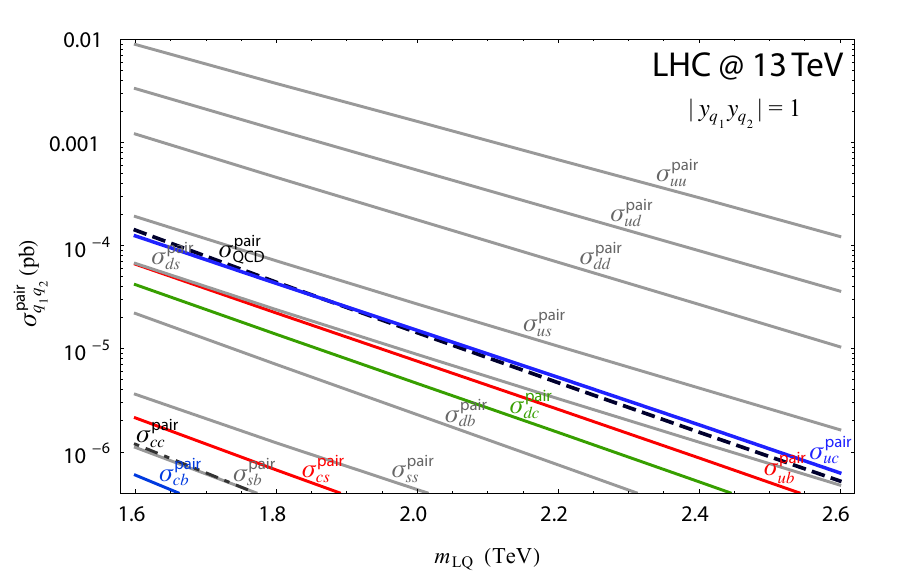}
\caption{Asymmetric leptoquark pair production cross sections $\sigma^\mathrm{pair}_{q_1 q_2}(y_{q_1},y_{q_2}, m_{\mathrm{LQ}})$ for quark-quark initial states, where $q_1=u,d,s,c$ and $q_2=u,d,s,c,b$.}\label{fig:C_S_qq}
\end{figure}

\begin{figure}[th!]\centering
\includegraphics[width=0.95\textwidth]{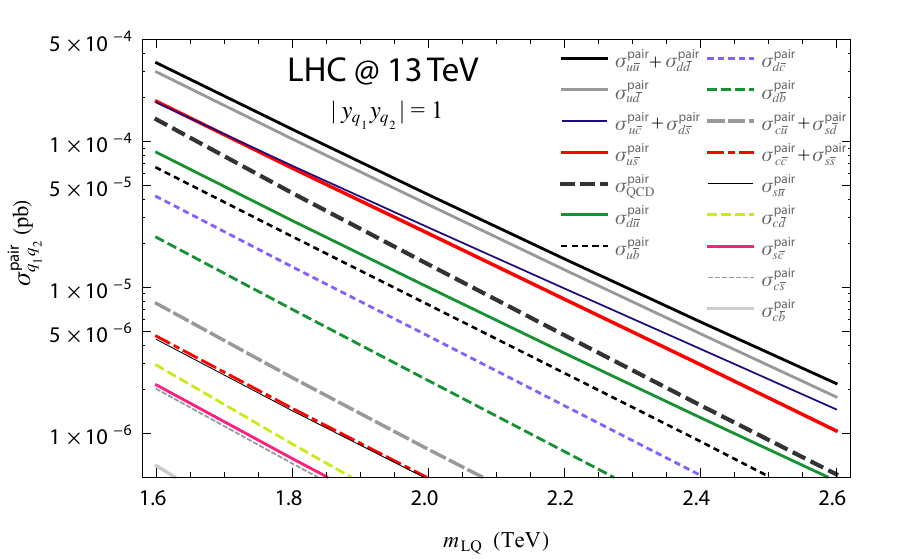}
\caption{Asymmetric leptoquark pair production cross sections $\sigma^\mathrm{pair}_{q_1 q_2}(y_{q_1},y_{q_2}, m_{\mathrm{LQ}})$ for quark-antiquark initial states, where $q_1=u,d,s,c$ and $q_2=\overline{u},\overline{d},\overline{s},\overline{c}, \overline{b}$.}\label{fig:C_S_qaq}
\end{figure}

The quark-quark initiated cross sections are given in Fig.~\ref{fig:C_S_qq} under the assumption that $|y_{q_1}y_{q_2}| =1$, where $q_1=u,d,s,c$ and $q_2=u,d,s,c,b$, while the quark-antiquark initiated cross sections are given in Fig.~\ref{fig:C_S_qaq} under the same assumption that $|y_{q_1}y_{q_2}| =1$, but, this time around, with $q_1=u,d,s,c$ and $q_2=\overline{u},\overline{d},\overline{s},\overline{c},\overline{b}$. We also present in Figs.~\ref{fig:C_S_qq} and \ref{fig:C_S_qaq}, for comparison purposes, conventional scalar leptoquark pair production cross section at the LHC that is evaluated under the assumption that the leptoquark Yukawa couplings are negligible but still large enough to ensure prompt leptoquark decay. This particular cross section is simply denoted with $\sigma^\mathrm{pair}_\mathrm{QCD}(m_{\mathrm{LQ}})$ to stress that it is purely QCD induced and it is represented by a thick dashed black curve in both Figs.~\ref{fig:C_S_qq} and~\ref{fig:C_S_qaq}.

The asymmetric leptoquark pair production cross sections of Figs.~\ref{fig:C_S_qq} and~\ref{fig:C_S_qaq} are extracted from the new physics scenarios of Fig.~\ref{fig:LQ_COMBINATIONS}, where all of them are generated by the $t$-channel processes of
Fig.~\ref{diag-type}. These scenarios are implemented using {\rmfamily\scshape FeynRules}~\cite{Alloul:2013bka} and subsequently imported in {\rmfamily\scshape MadGraph5\_aMC@NLO} framework~\cite{Alwall:2014hca} to produce numerical results for $m_\mathrm{LQ}$ values between 1.6\,TeV and 2.6\,TeV. We exclusively use the {\tt nn23lo1} PDF set~\cite{NNPDF:2014otw} to generate leading order cross sections for the center-of-mass energy of proton-proton collisions set at 13\,TeV, where the factorisation ($\mu_F$) and renormalization ($\mu_R$) scales are taken to be $\mu_F=\mu_R= m_\mathrm{LQ}/2$. Note that we only quote central values for all cross sections as we are solely interested in relative strengths of various potential contributions. 

One can observe from Fig.~\ref{fig:C_S_qq} that the quark-quark initiated asymmetric pair production cross sections of mass-degenerate scalar leptoquarks $\mathrm{LQ}_1$ and $\mathrm{LQ}_2$, i.e., when $\Delta F =|F(\mathrm{LQ}_1)|-|F(\mathrm{LQ}_2)|=\pm 2$ and $m_{\mathrm{LQ}_1}=m_{\mathrm{LQ}_2}=m_\mathrm{LQ}$, can be comparable to or be even substantially larger than the QCD driven leptoquark pair production cross section at the LHC if at least one of the leptoquarks couples to a valence quark and the product of relevant Yukawa cuplings is of order one. For example, $\sigma^\mathrm{pair}_{u u}(y_{q_1},y_{q_2}, m_{\mathrm{LQ}})$ for $|y_{q_1} y_{q_2}|=1$ is approximately two orders of magnitude larger than $\sigma^\mathrm{pair}_\mathrm{QCD}(m_{\mathrm{LQ}})$. In all other instances, i.e., for $q_1,q_2=s,c,b$, the cross sections for asymmetric pair production are, at best, a tiny correction of the QCD driven one, again, for order one Yukawa coupling strengths. 
In the case of the quark-antiquark initiated asymmetric pair production cross sections the only truly relevant scenarios, once again, are those where the initial quark is a valence quark. This is nicely illustrated in Fig.~\ref{fig:C_S_qaq} with the direct comparison between the QCD cross section rendered with a black thick dashed curve and the quark-antiquark induced cross sections $\sigma^\mathrm{pair}_{q_1 q_2}(y_{q_1},y_{q_2}, m_{\mathrm{LQ}})$, where $q_1=u,d,s,c$ and $q_2=\overline{u},\overline{d},\overline{s},\overline{c},\overline{b}$ and $|y_{q_1} y_{q_2}|=1$. The reason why we opted to plot combinations $\sigma^\mathrm{pair}_{u\overline{u}}+\sigma^\mathrm{pair}_{d\overline{d}}$, $\sigma^\mathrm{pair}_{u\overline{c}}+\sigma^\mathrm{pair}_{d\overline{s}}$, $\sigma^\mathrm{pair}_{c\overline{u}}+\sigma^\mathrm{pair}_{s\overline{d}}$ and $\sigma^\mathrm{pair}_{c\overline{c}}+\sigma^\mathrm{pair}_{s\overline{s}}$ in Fig.~\ref{fig:C_S_qaq}, instead of individual cross sections, will be elaborated on in Sec.~\ref{sec:S1S3}. 

Figs.~\ref{fig:C_S_qq} and~\ref{fig:C_S_qaq} demonstrate that it is entirely possible to have substantial cross sections for the asymmetric leptoquark pair production even when one of the leptoquarks couples weakly to the first generation of quarks whereas the other leptoquark couples strongly to the second or third generation of quarks as long as they both couple to a lepton of the same flavor and chirality. Of course, quark-quark initiated processes of asymmetric leptoquark pair production, i.e., when $\Delta F=|F(\mathrm{LQ}_1)|-|F(\mathrm{LQ}_2)|=\pm 2$, are potentially much more relevant at the LHC whereas quark-antiquark initiated processes are naturally enhanced at the Tevatron like machines. Note that the QCD cross section drops faster than the asymmetric cross sections as the mass of leptoquarks is increased. This is due to the fact that the gluon-gluon initiated processes start to be subdominant with respect to the processes initiated by quarks once one goes towards the large leptoquark mass limit.

Before we give an explicit example of the potential importance of the asymmetric scalar leptoquark pair production mechanism we want to address one subtlety associated with the conventional leptoquark pair production that has not been discussed in the literature before.

Conventional leptoquark pair production amplitude, when the leptoquarks comprising a pair are charge conjugates of each other, has two distinct contributions at the leading order. The first one is of purely QCD nature whereas the second one exhibits quadratic dependence on the leptoquark Yukawa coupling $y_{q}$. For a single scalar leptoquark $\mathrm{LQ}$ that couples to a quark $q$ and any lepton $l$ with Yukawa coupling $y_{q}$ the conventional leptoquark pair production cross section can thus be written as
\begin{equation}
\label{eq:cross}
    \sigma^\mathrm{pair}_{q \overline{q}}(y_{q}, m_\mathrm{LQ})=\sigma^\mathrm{pair}_\mathrm{QCD}( m_\mathrm{LQ})+a^\mathrm{interference}_{q \overline{q}}( m_\mathrm{LQ})y_{q}^2+\sigma^\mathrm{pair}_{q \overline{q}}(y_{q},y_{q}, m_\mathrm{LQ}),
\end{equation}
where $\sigma^\mathrm{pair}_\mathrm{QCD}( m_\mathrm{LQ})$  and $\sigma^\mathrm{pair}_{q \overline{q}}(y_{q},y_{q}, m_\mathrm{LQ})$ have been featured before and we assume, for consistency, that $y_{q}$ is real. If $y_{q}$ is small, the cross section depends solely on the leptoquark mass $m_\mathrm{LQ}$ and the particularities associated with the hadron machine itself and it is given by $\sigma^\mathrm{pair}_\mathrm{QCD}( m_\mathrm{LQ})$. In fact, $\sigma^\mathrm{pair}_\mathrm{QCD}( m_\mathrm{LQ})$ has been known analytically at the next-to-leading order in QCD for a long time~\cite{Kramer:2004df}.

The last term in Eq.~\eqref{eq:cross} corresponds to a $t$-channel exchange of a lepton $l$ with the $q \overline{q}$ pair in the initial state. Again, it does not depend on the type of lepton that the leptoquark couples to and the relevant cross sections are already introduced in Fig.~\ref{fig:C_S_qaq} for $y_{q}=1$. Finally, there is the interference term $a^\mathrm{interference}_{q \overline{q}}( m_\mathrm{LQ}) y_{q}^2$ that turns out to always be negative. There is thus a dip in the pair production cross section below the $\sigma^\mathrm{pair}_\mathrm{QCD}( m_\mathrm{LQ})$ value as Yukawa coupling is increased before $y_q$ becomes sufficiently large to make the third term in Eq.~\eqref{eq:cross} that is of quartic nature in terms of $y_q$ to start to dominate over the interference term that is of quadratic nature in $y_q$. We note, that the automated inclusion of the $t$-channel term at the next-to-leading order in QCD has been recently introduced in the literature~\cite{Borschensky:2021hbo}. 

The things, though, can change with regard to interference effect if a scalar leptoquark $\mathrm{LQ}$ couples to a quark $q$ and any lepton $l$ with Yukawa coupling $y_{q}$ and another quark $q^\prime$ and the same lepton with Yukawa coupling $y_{q^\prime}$. There will then exist four interference terms $a^\mathrm{interference}_{q \overline{q}}( m_\mathrm{LQ}) y_{q}^2$, $a^\mathrm{interference}_{q^\prime \overline{q^\prime}}( m_\mathrm{LQ}) y_{q^\prime}^2$, $a^\mathrm{interference}_{q \overline{q^\prime}}( m_\mathrm{LQ}) y_{q} y_{q^\prime}$, and $a^\mathrm{interference}_{q^\prime \overline{q}}( m_\mathrm{LQ}) y_{q} y_{q^\prime}$, where the last two can obviously exhibit constructive interference if $y_{q}$ and $y_{q^\prime}$ differ in sign. In fact, it might be even possible for both $a^\mathrm{interference}_{q \overline{q}}( m_\mathrm{LQ}) y_{q}^2$ and $a^\mathrm{interference}_{q^\prime \overline{q^\prime}}( m_\mathrm{LQ}) y_{q^\prime}^2$ to be less relevant than either $a^\mathrm{interference}_{q \overline{q^\prime}}( m_\mathrm{LQ}) y_{q} y_{q^\prime}$ or $a^\mathrm{interference}_{q^\prime \overline{q}}( m_\mathrm{LQ}) y_{q} y_{q^\prime}$. The point we want to make here is that the conventional pair production of leptoquarks might be sensitive not only to Yukawa coupling strengths but also to the relative sign between relevant Yukawa couplings even when these couplings are taken to be real.

With these preliminary considerations out of the way we now turn towards quantitative analysis of the asymmetric pair production mechanism within several concrete scenarios of new physics.

\subsection{Case studies}

Our primary aim is to advocate importance of inclusion of the asymmetric pair production mechanism in a quantitative determination of the viable leptoquark parameter space if and when appropriate. To that end, we discuss five different leptoquark extensions of the SM and derive, for several particular realisations of these extensions, accurate limits using two specific  experimental searches. More specifically, we recast the ATLAS Collaboration analysis~\cite{ATLAS:2020dsk} of the leptoquark pair production searches via $pp\rightarrow \mathrm{LQ} \overline{\mathrm{LQ}} \rightarrow jjee$ and $pp\rightarrow \mathrm{LQ} \overline{\mathrm{LQ}} \rightarrow jj \mu \mu$ processes, where $j$ is taken to generically represents a light jet, i.e., $j=u,\overline{u},d,\overline{d},s,\overline{s}$, while it is implicitly understood that both $ee$ and $\mu\mu$ stand for oppositely charged lepton pairs. All five scenarios provide a setting for pedagogical illustration of various phenomenological intricacies associated with the leptoquark pair production signatures. 

First of these five scenarios involves a presence of a single scalar leptoquark $S_1$. The second scenario of new physics is an $R_2$ extension of the SM, where $R_2$ multiplet comprises two states, i.e., $R_2^{+5/3}$ and $R_2^{-2/3}$. Third scenario extends the SM particle content with both $S_1$ and $R_2$ while forth scenario concerns addition of an $S_3$ leptoquark multiplet to the SM particle content, where $S_3$ contains scalars $S_3^{+4/3}$, $S_3^{+1/3}$, and $S_3^{-2/3}$. Fifth scenario regards simultaneous extension of the SM with both $S_1$ and $S_3$. We will assume that all these leptoquarks exclusively couple to either electrons or muons and the first generation quarks, to simplify discussion, where, for the electron coupling case, we also produce in Sec.~\ref{sec:APV} the latest APV search limits on the leptoquark parameter spaces.

Relevant parts of the $S_1$ lagrangian, for our study, are
\begin{align}
\nonumber
\mathcal{L}_{S_1} = &+y^{LL}_{1\,ij}\bar{Q}_{L}^{C\,i,a} S_{1} \epsilon^{ab}L_{L}^{j,b}+y^{RR}_{1\,ij}\bar{u}_{R}^{C\,i} S_{1} e_{R}^{j}+\textrm{h.c.}\\
\label{eq:main_S_1}
=&-(y^{LL}_1 U)_{ij} \bar{d}_{L}^{C\,i} \nu_{L}^{j} S_{1}^{+1/3}+(V^* y^{LL}_1)_{ij}\bar{u}_{L}^{C\,i} e_{L}^{j} S_{1}^{+1/3}+y^{RR}_{1\,ij}\bar{u}_{R}^{C\,i} e_{R}^{j} S_{1}^{+1/3}+\textrm{h.c.},
\end{align}
where $a,b(=1,2)$ are $SU(2)$ indices, $V$ is a Cabibbo--Kobayashi--Maskawa (CKM) mixing matrix, and $U$ represents a Pontecorvo--Maki--Nakagawa--Sakata (PMNS) unitary mixing matrix. We set the CKM matrix to be an identity matrix whereas the exact form of the PMNS matrix is irrelevant for our considerations as long as it resides entirely in the neutrino sector. Note that the CKM matrix, in our convention, is in the up-type quark sector. We will address validity of our assumption that the off-diagonal CKM matrix elements can be neglected and whether the exact placement of the CKM matrix is of any importance.

Pertinent parts of the $R_2$ lagrangian are
\begin{align}
\nonumber
\mathcal{L}_{R_2} =&-y^{RL}_{2\,ij}\bar{u}_{R}^{i} R_{2}^{a}\epsilon^{ab}L_{L}^{j,b}+y^{LR}_{2\,ij}\bar{e}_{R}^{i} R_{2}^{a\,*}Q_{L}^{j,a} +\textrm{h.c.}\\
\nonumber
= & - y^{RL}_{2\,ij} \bar{u}^i_R e^j_L R_2^{+5/3} + (y^{RL}_2
  U)_{ij} \bar{u}^i_R
  \nu^j_L R_2^{+2/3}+\\
 \label{eq:main_R_2}
 &+(y^{LR}_2 V^\dagger)_{ij} \bar{e}^i_R 
  u^j_L R_2^{-5/3} +y^{LR}_{2\,ij} \bar{e}^i_R d^j_L R_2^{-2/3} + \textrm{h.c.}.
\end{align}
One can note that all unitary transformations of the right-chiral fermions can be completely absorbed, for both the $S_1$ and $R_2$ scenarios, into associated Yukawa coupling matrices. We accordingly take all unitary transformations of right-chiral quarks and charged leptons to be unphysical in our study. 

The $S_3$ Lagrangian, in our notation, is
\begin{align}
\nonumber
\mathcal{L}_{S_3} &= y^{LL}_{3\,ij}\bar{Q}_{L}^{C\,i,a} \epsilon^{ab} (\tau^k S^k_{3})^{bc} L_{L}^{j,c} +\mathrm{h.c.}\,\\
\nonumber
&=-(y^{LL}_{3}U)_{ij}\bar{d}_{L}^{C\,i} \nu_{L}^{j} S^{+1/3}_{3}-(V^* y^{LL}_{3})_{ij}\bar{u}_{L}^{C\,i} e_{L}^{j} S^{+1/3}_{3}+\\
\label{eq:main_S_3}
&+\sqrt{2} (V^* y^{LL}_{3}U)_{ij}\bar{u}_{L}^{C\,i} \nu_{L}^{j} S^{-2/3}_{3}-\sqrt{2} y^{LL}_{3\,ij}\bar{d}_{L}^{C\,i} e_{L}^{j} S^{+4/3}_{3} +\text{h.c.},
\end{align}
where $\tau^k$, $k=1,2,3$, are Pauli matrices and we define $S^{+4/3}_3=(S^1_3-i S^2_3)/\sqrt{2}$, $S^{+1/3}_3=S^3_3$, and $S^{-2/3}_3=(S^1_3+i S^2_3)/\sqrt{2}$ to be electric charge eigenstates.

Again, in all of these scenarios we will always assume a presence of a single non-zero Yukawa coupling to either electron or muon and the first generation quarks for each of these leptoquark multiplets, if and when they are featured, in order to simplify discussion.

\subsubsection{Case study: $S_1(\overline{\mathbf{3}},\mathbf{1},1/3)$}
\label{sec:S1}

Let us first address the $S_1$ scenario. 

\begin{figure}[t!]\centering
\includegraphics[width=0.95\textwidth]{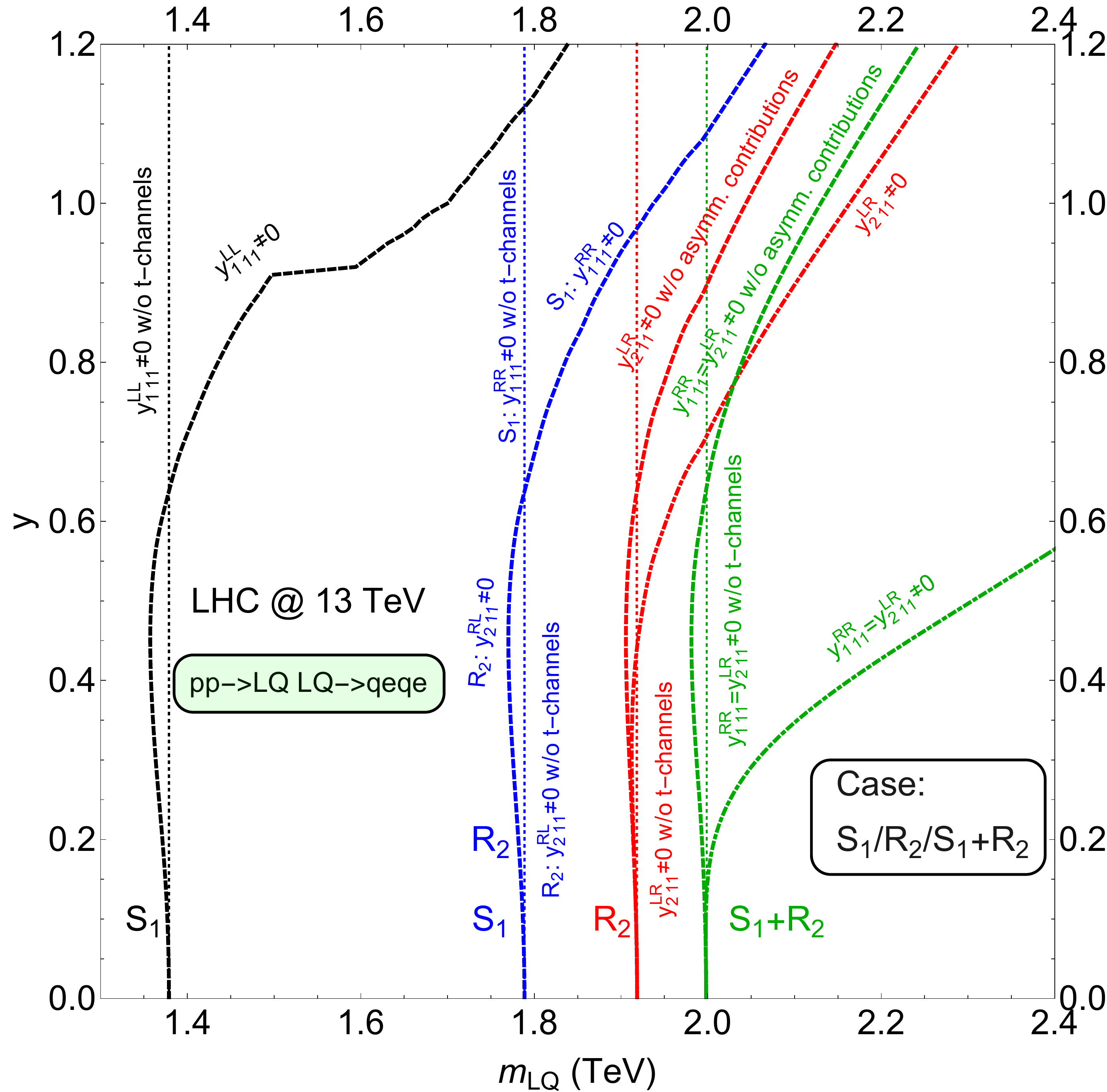}
\caption{The leptoquark parameter space limits for the $S_1$, $R_2$, and $S_1$+$R_2$ scenarios extracted from the $pp\rightarrow \mathrm{LQ} \overline{\mathrm{LQ}} \rightarrow jjee$ process search~\cite{ATLAS:2020dsk} performed at 13\,TeV center-of-mass energy of proton-proton collisions at the LHC, using an integrated luminosity of 139\,fb$^{-1}$. See text for more details.}\label{fig:LIMITS1}
\end{figure}

\begin{figure}[t!]\centering
\includegraphics[width=0.95\textwidth]{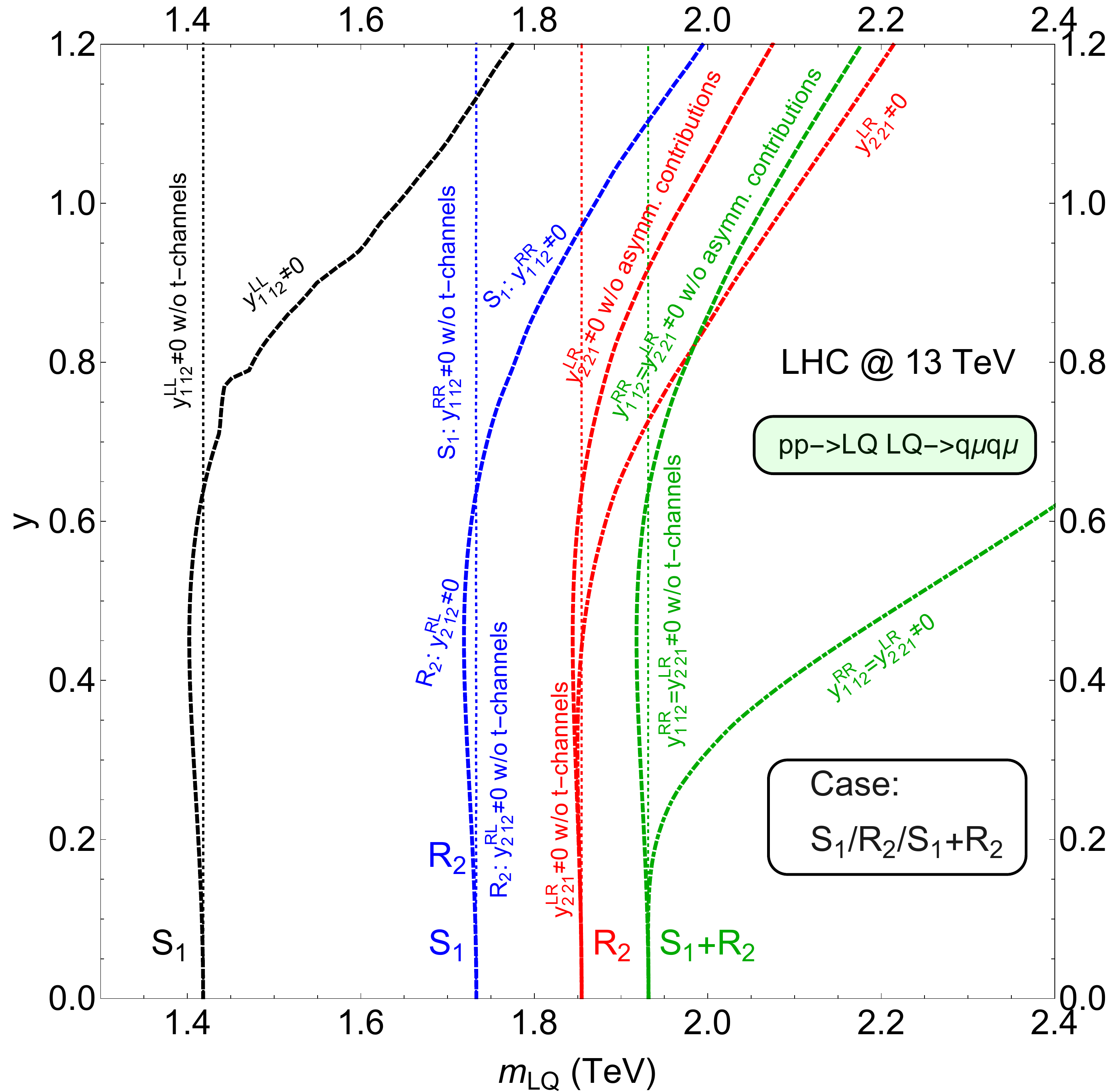}
\caption{The leptoquark parameter space limits for the $S_1$, $R_2$, and $S_1$+$R_2$ scenarios extracted from the $pp\rightarrow \mathrm{LQ} \overline{\mathrm{LQ}} \rightarrow jj\mu\mu$ process search~\cite{ATLAS:2020dsk} performed at 13\,TeV center-of-mass energy of proton-proton collisions at the LHC, using an integrated luminosity of 139\,fb$^{-1}$. See text for more details.}\label{fig:LIMITS3}
\end{figure}

\begin{itemize}
\item If we assume that $y^{LL}_{1\,11} \equiv y$ is the only non-zero Yukawa coupling present in Eq.~\eqref{eq:main_S_1}, we have that the branching ratios for the $S_1$ decays are $B(S_{1}^{\pm 1/3} \rightarrow j \nu)=1/2$ and $B(S_{1}^{\pm 1/3} \rightarrow j e)=1/2$. A recast of the ATLAS Collaboration analysis~\cite{ATLAS:2020dsk} of the leptoquark pair production search via $pp\rightarrow \mathrm{LQ} \overline{\mathrm{LQ}} \rightarrow jjee$ process at 13\,TeV center-of-mass energy of proton-proton collisions, using an integrated luminosity of 139\,fb$^{-1}$, then yields a limit on the mass of $S_1$ leptoquark, as a function of $y \equiv y^{LL}_{1\,11}$, which is rendered with a thick dashed black curve in Fig.~\ref{fig:LIMITS1}. The exclusion region is to the left of that curve and it is based on the ATLAS Collaboration observed 95\% C.L.\ limit. The Yukawa dependent limit we present in Fig.~\ref{fig:LIMITS1}, for small values of $y^{LL}_{1\,11} \equiv y$, needs to agree with the outcome of the ATLAS Collaboration analysis when $B(S_{1}^{\pm 1/3} \rightarrow j e)=1/2$, i.e. $m_\mathrm{LQ} \geq 1380$\,GeV, which is based on the next-to-leading order cross section in QCD calculation~\cite{ATLAS:2020dsk}. We accordingly rescale our leading order simulation when presenting the limits in Fig.~\ref{fig:LIMITS1} and note that the cross section obtained in that way indeed corresponds to the next-to-leading order cross section in QCD as given in Ref.~\cite{Dorsner:2018ynv}. We also plot in Fig.~\ref{fig:LIMITS1} the leptoquark parameter constraint with a vertical thin dashed black line if one would use $\sigma_\mathrm{QCD}^\mathrm{pair}(m_\mathrm{LQ})$ instead of the Yukawa dependant cross section, for this particular branching fraction scenario. That vertical line is additionally marked with ``w/o $t$-channels'' to stress exclusion of the $t$-channel lepton exchange diagrams during evaluation of $\sigma_\mathrm{QCD}^\mathrm{pair}(m_\mathrm{LQ})$.

We note two subtleties with regard to the $y^{LL}_{1\,11}  \equiv y \neq 0$ case. First, there are two $t$-channel contributions towards the $S_1$ pair production that need to be included in this analysis. One contribution is due to the first term in the second line of Eq.~\eqref{eq:main_S_1} and it is $d \overline{d}$ initiated. The other contribution is $u \overline{u}$ initiated and it is due to the second term in the second line of Eq.~\eqref{eq:main_S_1}. Another subtlety concerns the CKM mixing matrix placement. Namely, if the CKM matrix is taken to be in the up-type quark sector it would induce coupling between $S_1$, a charm quark, and an electron through the second term in the second line of Eq.~\eqref{eq:main_S_1}. This would primarily impact the branching ratio $B(S_{1}^{\pm 1/3} \rightarrow j e)$ by reducing it to 80\% of its initial value and would also introduce $B(S_{1}^{\pm 1/3} \rightarrow c e)$ at the level of 10\%. These changes in branching fractions would consequentially impact interpretation of the ATLAS Collaboration analysis~\cite{ATLAS:2020dsk} that can distinguish between light jets and, for example, a $c$-quark induced jet. The bounds on the $S_1$ parameter space would accordingly shift to the left in Fig.~\ref{fig:LIMITS1}. The placement of the CKM mixing matrix in the down-type quark sector, on the other hand, would not produce any such shift.

\item If we take $y^{LL}_{1\,12} \equiv y \neq 0$ in Eq.~\eqref{eq:main_S_1}, the branching ratios for the $S_1$ decays read $B(S_{1}^{\pm 1/3} \rightarrow j \nu)=1/2$ and $B(S_{1}^{\pm 1/3} \rightarrow j \mu)=1/2$. A recast of the ATLAS Collaboration analysis~\cite{ATLAS:2020dsk} of the leptoquark pair production search via $pp\rightarrow \mathrm{LQ} \overline{\mathrm{LQ}} \rightarrow jj \mu \mu$ process then yields a limit on the mass of $S_1$ leptoquark, as a function of $y  \equiv y^{LL}_{1\,12}$, which is rendered with a thick dashed black curve in Fig.~\ref{fig:LIMITS3}. The limit we present in Fig.~\ref{fig:LIMITS3}, for small values of $y^{LL}_{1\,12} \equiv y$, corresponds to the outcome of the ATLAS Collaboration analysis when $B(S_{1}^{\pm 1/3} \rightarrow j \mu)=1/2$, i.e. $m_\mathrm{LQ} \geq 1420$\,GeV, that is shown as a vertical thin dashed black line in Fig.~\ref{fig:LIMITS3}.

\item If we take that $y^{RR}_{1\,11} \equiv y$ is the only non-zero Yukawa coupling in Eq.~\eqref{eq:main_S_1}, we get that $B(S_{1}^{\pm 1/3} \rightarrow j e)=1$ and the correct interpretation of the ATLAS Collaboration results~\cite{ATLAS:2020dsk} would correspond to a bound rendered with a thick dashed blue curve in Fig.~\ref{fig:LIMITS1}. This bound, for small values of $y^{RR}_{1\,11}  \equiv y$, yields $m_\mathrm{LQ} \geq 1790$\,GeV~\cite{ATLAS:2020dsk} and thus coincides with the constraint presented with a vertical thin dashed blue line that is generated if one were to use $\sigma_\mathrm{QCD}^\mathrm{pair}(m_\mathrm{LQ})$ instead of the more appropriate $\sigma_{u \overline{u}}^\mathrm{pair}(y,m_\mathrm{LQ})$ to interpret the ATLAS Collaboration analysis.

\item If we take that $y^{RR}_{1\,12} \equiv y \neq 0$, a recast of the ATLAS Collaboration results~\cite{ATLAS:2020dsk} on the $pp\rightarrow \mathrm{LQ} \overline{\mathrm{LQ}} \rightarrow jj \mu \mu$ process search yields a bound rendered with a thick dashed blue curve in Fig.~\ref{fig:LIMITS3}. This bound, for small values of $y^{RR}_{1\,12}  \equiv y$, reads $m_\mathrm{LQ} \geq 1730$\,GeV~\cite{ATLAS:2020dsk} and is given with a vertical thin dashed blue line in Fig.~\ref{fig:LIMITS3}.
\end{itemize}

Note that the exclusion regions, in all four cases, feature negative interference effects, as discussed in connection to Eq.~\eqref{eq:cross}, for intermediate values of Yukawa couplings $y^{LL}_{1\,11} \equiv y$, $y^{LL}_{1\,12}  \equiv y$, $y^{RR}_{1\,11}  \equiv y$, and $y^{RR}_{1\,12}  \equiv y$.

\subsubsection{Case study: $R_2(\mathbf{3},\mathbf{2},7/6)$}
\label{sec:R2}

We consider, in what follows, scenarios when we switch on, individually, $y^{RL}_{2\,11}$, $y^{RL}_{2\,12}$, $y^{LR}_{2\,11}$, and $y^{LR}_{2\,12}$ of Eq.~\eqref{eq:main_R_2} while all other Yukawa matrix elements are taken to be negligible.

\begin{figure}[th!]
\centering
\includegraphics[width=1\textwidth]{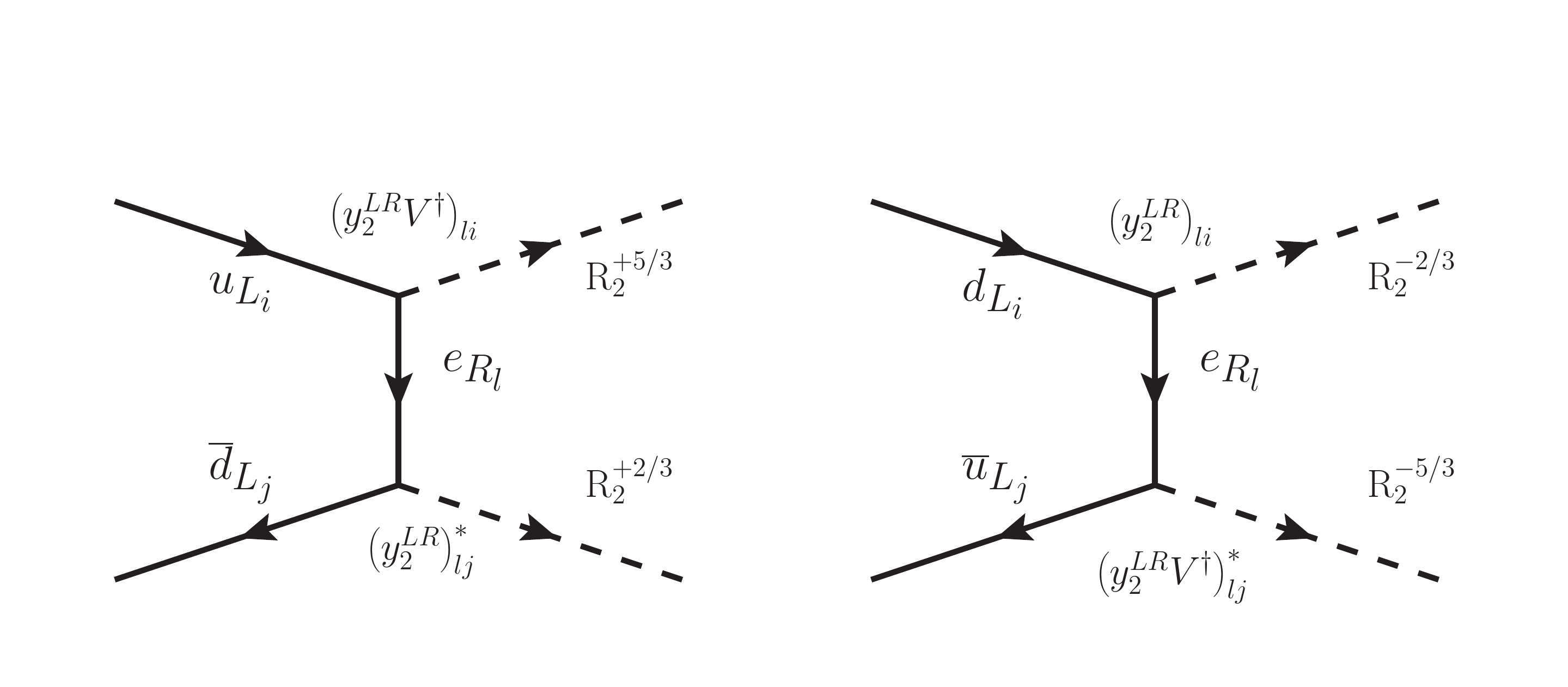}
\caption{Asymmetric pair production for the case of $R_2$ leptoquark.}\label{fig:R2}
\end{figure}

\begin{itemize}
\item If we turn on Yukawa coupling $y^{RL}_{2\,11} \equiv y$ in Eq.~\eqref{eq:main_R_2}, we have that $B(R_{2}^{\pm 5/3} \rightarrow j e)=1$ and $B(R_{2}^{\pm 2/3} \rightarrow j \nu)=1$. Since the members of the $R_2$ multiplet need to be mass-degenerate for all practical purposes, the limit on the $R_{2}^{\pm 5/3}$ parameter space, as extracted from the ATLAS Collaboration pair production analysis~\cite{ATLAS:2020dsk}, should also be applicable to $R_{2}^{\pm 2/3}$ and vice versa. If we furthermore take into account the fact that the experimental limit on $pp\rightarrow R_{2}^{+5/3} R_{2}^{-5/3} \rightarrow jj ee$ is certainly more relevant than the limit that could be extracted from $pp\rightarrow R_{2}^{+2/3} R_{2}^{-2/3} \rightarrow jj \nu \nu$, the constraint on the viable $R_{2}^{\pm 5/3}$ and $R_{2}^{\pm 2/3}$ parameter spaces is given with a thick dashed blue curve in Fig.~\ref{fig:LIMITS1}. Note that this particular limit on the $R_{2}^{\pm 5/3}$ and $R_{2}^{\pm 2/3}$ parameter spaces, when $y^{RL}_{2\,11} \equiv y \neq 0$, is the same as for $S_1$ leptoquark when $y^{RR}_{1\,11} \equiv y \neq 0$. 

\item If we assume that $y^{RL}_{2\,12}  \equiv y \neq 0$ in Eq.~\eqref{eq:main_R_2}, we have that $B(R_{2}^{\pm 5/3} \rightarrow j \mu)=1$ and $B(R_{2}^{\pm 2/3} \rightarrow j \nu)=1$. In analogy to the previous case the reinterpretation of the ATLAS Collaboration pair production analysis~\cite{ATLAS:2020dsk} of the $pp\rightarrow \mathrm{LQ} \overline{\mathrm{LQ}} \rightarrow jj \mu \mu$ process search yields a constraint on the viable $R_{2}^{\pm 5/3}$ and $R_{2}^{\pm 2/3}$ parameter spaces that is given with a thick dashed blue curve in Fig.~\ref{fig:LIMITS3}. A vertical thin dashed blue line in Fig.~\ref{fig:LIMITS3} represents a limit that is based solely on the QCD cross section and corresponds to $m_\mathrm{LQ} \geq 1730$\,GeV~\cite{ATLAS:2020dsk}. The limits on the $R_{2}^{\pm 5/3}$ and $R_{2}^{\pm 2/3}$ parameter spaces, when $y^{RL}_{2\,12} \equiv y \neq 0$, are the same as for $S_1$ leptoquark when $y^{RR}_{1\,12} \equiv y \neq 0$.

\item If we set $y^{LR}_{2\,11} \equiv y \neq 0$, we have that $B(R_{2}^{\pm 5/3} \rightarrow j e)=1$ and $B(R_{2}^{\pm 2/3} \rightarrow j e)=1$. Since the pair productions of both components of $R_2$ produce the same final state, i.e., $pp\rightarrow R_{2}^{+5/3} R_{2}^{-5/3} \rightarrow jj ee$ and $pp\rightarrow R_{2}^{+2/3} R_{2}^{-2/3} \rightarrow jj ee$, we need to take that into account. Naive combination of these two processes, i.e., based purely on the $\sigma_\mathrm{QCD}^\mathrm{pair}(m_\mathrm{LQ})$ value, results in a bound given by a vertical thin dashed red line in Fig.~\ref{fig:LIMITS1} and yields $m_\mathrm{LQ} \geq 1920$\,GeV. If we furthermore include the Yukawa dependence of the cross sections to pair produce both components of $R_2$ multiplet, we obtain a limit rendered in a thick dashed red curve in Fig.~\ref{fig:LIMITS1}. It should be noted that the generation of the thick dashed red curve denoted with ``w/o asymm.\ contributions'' calls for separate evaluation of cross sections for both $pp\rightarrow R_{2}^{+5/3} R_{2}^{-5/3} (\rightarrow jj ee)$ and $pp\rightarrow R_{2}^{+2/3} R_{2}^{-2/3} (\rightarrow jj ee)$ and their subsequent addition. Since $R_{2}^{-5/3}$ couples to the up quark while $R_{2}^{-2/3}$ couples to the down quark, these two cross sections, as functions of $y^{LR}_{2\,11} \equiv y$, are clearly not identical.

Note, however, that simple addition of cross sections to produce $R_{2}^{+5/3} R_{2}^{-5/3}$ and $R_{2}^{+2/3} R_{2}^{-2/3}$ pairs does not account for the asymmetric pair production mechanism effects that we want to advocate. To take into account asymmetric pair production we also need to include cross sections for $pp\rightarrow R_{2}^{+5/3} R_{2}^{-2/3} (\rightarrow jj ee)$ and $pp\rightarrow R_{2}^{-5/3} R_{2}^{+2/3} (\rightarrow jj ee)$. Relevant diagrams for these two processes are presented in Fig.~\ref{fig:R2}. The diagrams of Fig.~\ref{fig:R2} explicitly show that the two leptoquarks that are produced do not comprise a charge conjugate pair. These processes thus do not interfere, at the amplitude level, with the conventional pair production mechanisms even though they yield the exact same $jjee$ final state.

If we combine both the conventional and asymmetric pair production cross sections, and apply the constraints obtained by the ATLAS Collaboration on the $pp \rightarrow \mathrm{LQ} \overline{\mathrm{LQ}}\rightarrow jjee$ process~\cite{ATLAS:2020dsk}, we obtain a proper bound rendered with a thick dot-dashed red curve in Fig.~\ref{fig:LIMITS1}. The relevance of the asymmetric contribution is, in our view, self-evident.

We finally present, for completeness, the leading order cross sections for  $pp\rightarrow R_{2}^{+5/3} R_{2}^{-5/3}$, $pp\rightarrow R_{2}^{+2/3} R_{2}^{-2/3}$, $pp\rightarrow R_{2}^{+5/3} R_{2}^{-2/3}$, and $pp\rightarrow R_{2}^{-5/3} R_{2}^{+2/3}$ in Table~\ref{table:c_s_R2} as functions of $y \equiv y^{LR}_{2\,11}$ and $m_\mathrm{LQ}$. Note, again, that the cross sections for $pp\rightarrow R_{2}^{+5/3} R_{2}^{-5/3}$ and $pp\rightarrow R_{2}^{+2/3} R_{2}^{-2/3}$ behave differently with respect to change of $y^{LR}_{2\,11} \equiv y$.

\item If we set $y^{LR}_{2\,12} \equiv y \neq 0$, we have that $B(R_{2}^{\pm 5/3} \rightarrow j \mu)=1$ and $B(R_{2}^{\pm 2/3} \rightarrow j \mu)=1$. The limits presented in Fig.~\ref{fig:LIMITS3} in red with a vertical thin dashed line, a thick dashed curve, and a thick dot-dashed curve correspond to a QCD only limit, a conventional Yukawa coupling dependent limit, and a proper limit that includes asymmetric production effects, respectively. These limits converge to the same bound of $m_\mathrm{LQ} \geq 1850$\,GeV, as required by the observed 95\% C.L.\ limit of the ATLAS Collaboration search for the $pp \rightarrow \mathrm{LQ} \overline{\mathrm{LQ}}\rightarrow jj\mu\mu$ process~\cite{ATLAS:2020dsk}. 
\end{itemize}

\begin{table}
\begin{center}
\begin{tabular}{| c | c | c | c | c | c |}\hline
$m_\textrm{LQ}$ & $y$ & $\sigma_{R_2^{+2/3}R_2^{-2/3}}$ (fb) & $\sigma_{R_2^{+5/3}R_2^{-5/3}}$ (fb) & $\sigma_{R_2^{+5/3}R_2^{-2/3}}$ (fb) & $\sigma_{R_2^{-5/3}R_2^{+2/3}}$ (fb)\\\hline\hline
& $0.1$ & $0.141$ & $0.141$ & $2.98 \times 10^{-5}$ & $8.39 \times 10^{-6}$\\\cline{2-6}
1.6\,TeV & $0.5$ & $0.138$ & $0.132$ & $0.0187$ & $0.00523$\\\cline{2-6}
& $1.0$ & $0.201$ & $0.287$ & $0.298$ & $0.0837$\\\hline
\hline
& $0.1$ & $0.0143$ & $0.0143$ & $3.67 \times 10^{-6}$ & $9.93 \times 10^{-7}$\\\cline{2-6}
2.0\,TeV & $0.5$ & $0.0139$ & $0.0131$ & $0.00292$ & $0.000622$\\\cline{2-6}
& $1.0$ & $0.0210$ & $0.0331$ & $0.0367$ & $0.00996$\\\hline
\hline
& $0.1$ & $0.00157$ & $0.00156$ & $4.80 \times 10^{-7}$ & $1.28 \times 10^{-7}$\\\cline{2-6}
2.4\,TeV & $0.5$ & $0.00151$ & $0.00139$ & $2.99 \times 10^{-4}$ & $0.796 \times 10^{-4}$\\\cline{2-6}
& $1.0$ & $0.00239$ & $0.00414$ & $0.00482$ & $0.00128$\\
\hline\end{tabular}
\end{center}
\caption{The leading order cross sections for the leptoquark pair production for the $R_2$ scenario in the proton-proton collisions at 13\,TeV center-of-mass energy when the $R_2$ components of mass $m_\textrm{LQ}$ couple exclusively to a right-chiral leptons and the first generation quarks, as allowed by the SM gauge group, with the coupling strength $y\equiv y^{LR}_{2\,11}$.}
\label{table:c_s_R2}
\end{table}

\subsubsection{Case study: $S_1(\overline{\mathbf{3}},\mathbf{1},1/3)$+$R_2(\mathbf{3},\mathbf{2},7/6)$}
\label{sec:S1R2}

Since both $S_1$ and $R_2$ multiplets can couple to the SM leptons of both chiralities, there are four different scenarios to consider even if only one Yukawa coupling for each of these two multiplets is turned on at a given time. To avoid overburdening the reader with too many details and to drive our point of potential importance of the asymmetric pair production inclusion, we will investigate only one of these four possibilities for both the $jjee$ and $jj\mu\mu$ final state scenarios. 

\begin{itemize}
\item We first take that $y^{RR}_{1\,11} = y^{LR}_{2\,11} \equiv y \neq 0$ so that all three leptoquarks decay into the same final state, i.e., $B(S_{1}^{\pm 1/3} \rightarrow j e)= B(R_{2}^{\pm 5/3} \rightarrow j e)= B(R_{2}^{\pm 2/3} \rightarrow j e)=1$. If we furthermore take that the masses of $S_1$ and of the two charged components in $R_2$ are the same, we obtain that a pure QCD cross section generates bound given with a vertical thin  dashed green line whereas a simplistic addition of cross sections for processes $pp\rightarrow S_{1}^{+1/3} S_{1}^{-1/3} (\rightarrow jj ee)$, $pp\rightarrow R_{2}^{+5/3} R_{2}^{-5/3} (\rightarrow jj ee)$, and  $pp\rightarrow R_{2}^{+2/3} R_{2}^{-2/3} (\rightarrow jj ee)$ yields a bound given by a thick dashed green curve in Fig.~\ref{fig:LIMITS1}. These bounds are based on the observed 95\% C.L.\ limits, as given by the ATLAS Collaboration results on the $pp \rightarrow \mathrm{LQ} \overline{\mathrm{LQ}}\rightarrow jjee$ process search~\cite{ATLAS:2020dsk}, and yield $m_\mathrm{LQ} \geq 2000$\,GeV in the small Yukawa coupling limit. Since the ATLAS Collaboration analysis~\cite{ATLAS:2020dsk} provides results for the leptoquark masses up to 2\,TeV only, we conservatively assume that the observed limits above 2\,TeV would have the ATLAS Collaboration 2\,TeV level values.

If we finally include all six asymmetric contributions, i.e., $pp\rightarrow S_{1}^{\pm 1/3} R_{2}^{\pm 5/3} (\rightarrow jj ee)$, $pp\rightarrow S_{1}^{\pm 1/3} R_{2}^{\pm 2/3} (\rightarrow jj ee)$, and $pp\rightarrow R_{2}^{\pm 5/3} R_{2}^{\mp 2/3} (\rightarrow jj ee)$, we obtain a bound given by a thick dot-dashed green curve in Fig.~\ref{fig:LIMITS1}. Once again, the importance of inclusion of the asymmetric contribution is, in our view, self-evident. 

\item If we take that $y^{RR}_{1\,12} = y^{LR}_{2\,12} \equiv y \neq 0$ so that all three leptoquarks decay into muons and light jets, i.e., $B(S_{1}^{\pm 1/3} \rightarrow j \mu)= B(R_{2}^{\pm 5/3} \rightarrow j \mu)= B(R_{2}^{\pm 2/3} \rightarrow j \mu)=1$, we obtain the limits rendered in Fig.~\ref{fig:LIMITS3} in green. These limits are obtained under the assumption that all three leptoquarks are mass-degenerate and represent a recast of the ATLAS Collaboration search for the $pp \rightarrow \mathrm{LQ} \overline{\mathrm{LQ}}\rightarrow jj\mu\mu$ process~\cite{ATLAS:2020dsk}. A vertical thin dashed line is generated by a pure QCD cross section, a thick dashed curve is produced by a simple inclusion of the Yukawa dependent terms in the relevant cross sections while a thick dot-dashed curve corresponds to the correct inclusion of both the conventional and asymmetric contributions towards total cross section that yields the $jj\mu\mu$ final state. These three limits, rendered in green in Fig.~\ref{fig:LIMITS3}, converge to $m_\mathrm{LQ} \geq 1930$\,GeV for small values of $y^{RR}_{1\,12} = y^{LR}_{2\,12} \equiv y$.
\end{itemize}

\subsubsection{Case study: $S_3(\overline{\mathbf{3}},\mathbf{3},1/3)$}
\label{sec:S3}

We consider two particular scenarios for the $S_3$ case. One scenario is when $y^{LL}_{3\,11} \neq 0$ and the other one is when $y^{LL}_{3\,12} \neq 0$.

\begin{figure}[t!]\centering
\includegraphics[width=0.95\textwidth]{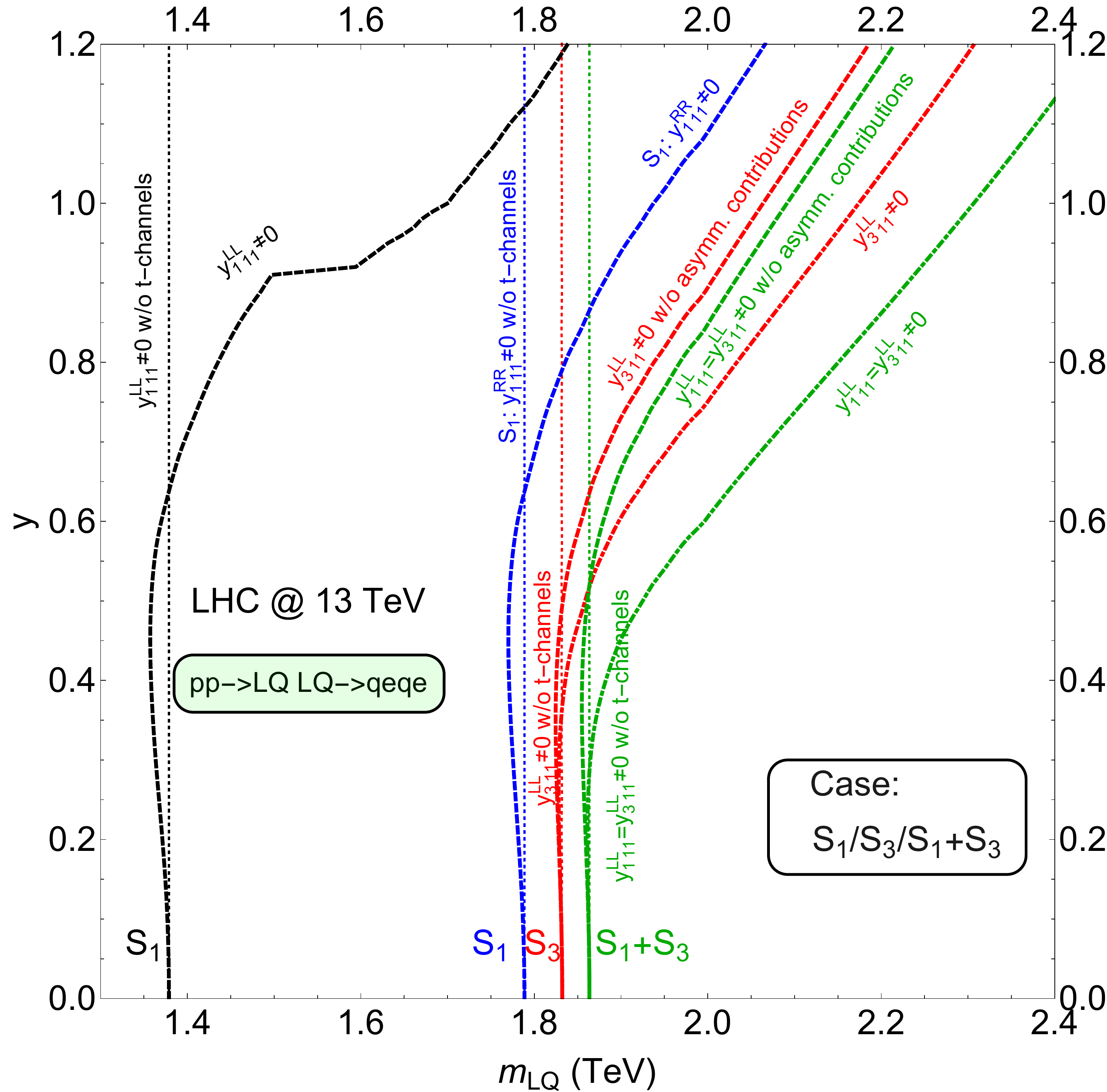}
\caption{The leptoquark parameter space limits for the $S_1$, $S_3$, and $S_1$+$S_3$ scenarios extracted from the $pp\rightarrow \mathrm{LQ} \overline{\mathrm{LQ}} \rightarrow jjee$ process search~\cite{ATLAS:2020dsk} performed at 13\,TeV center-of-mass energy of proton-proton collisions at the LHC, using an integrated luminosity of 139\,fb$^{-1}$. See text for more details.}\label{fig:LIMITS2}
\end{figure}

\begin{figure}[t!]\centering
\includegraphics[width=0.95\textwidth]{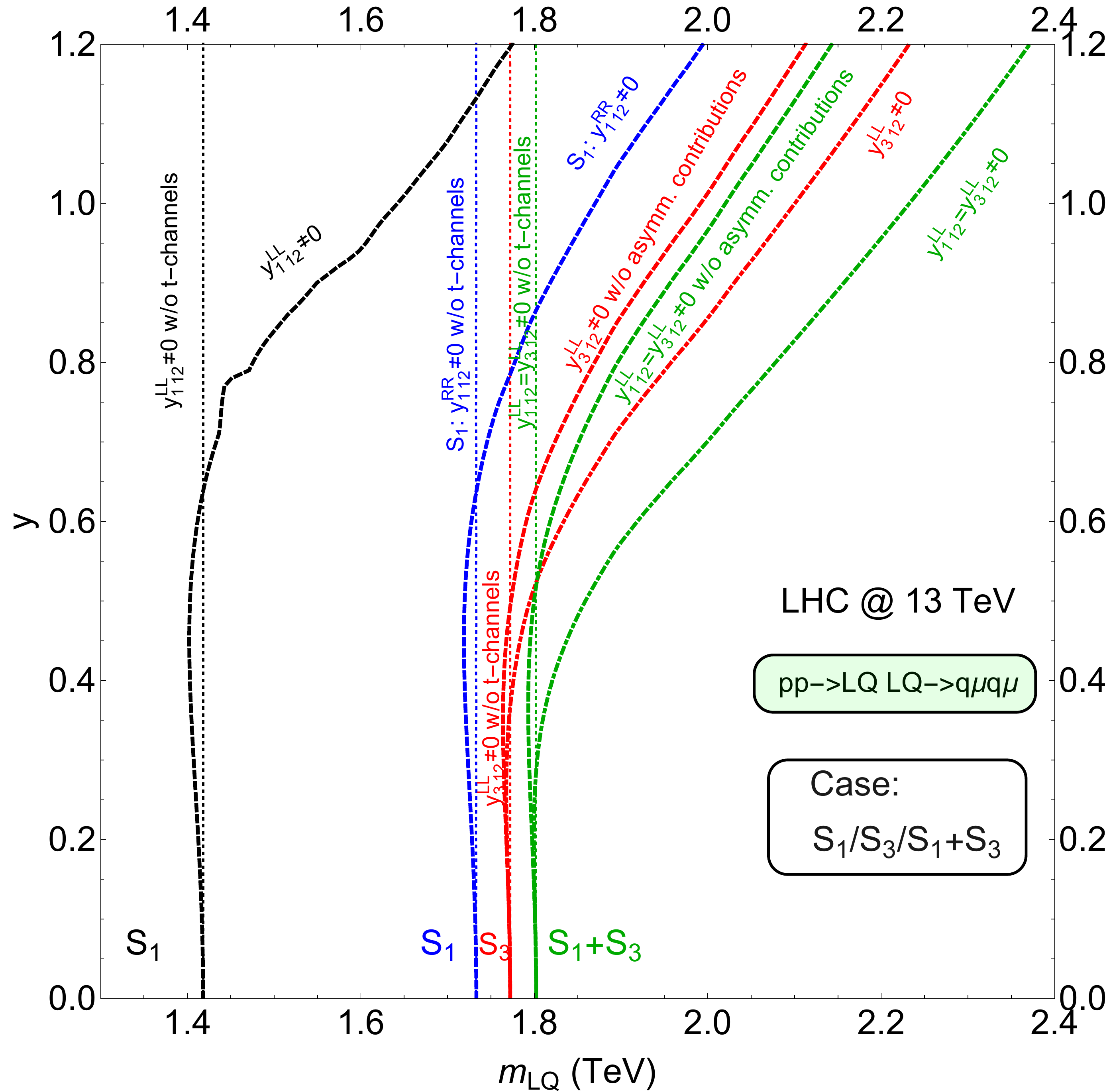}
\caption{The leptoquark parameter space limits for the $S_1$, $S_3$, and $S_1$+$S_3$ scenarios extracted from the $pp\rightarrow \mathrm{LQ} \overline{\mathrm{LQ}} \rightarrow jj\mu\mu$ process search~\cite{ATLAS:2020dsk} performed at 13\,TeV center-of-mass energy of proton-proton collisions at the LHC, using an integrated luminosity of 139\,fb$^{-1}$. See text for more details.}\label{fig:LIMITS4}
\end{figure}

\begin{itemize}
\item We assume that $y^{LL}_{3\,11} \equiv y$ of Eq.~\eqref{eq:main_S_3} is the only non-zero Yukawa coupling and take all three leptoquarks within $S_3$ multiplet to be degenerate in mass that we denote by $m_\mathrm{LQ}$. The branching fractions for the $S_3$ components, when $y^{LL}_{3\,11} \neq 0$, are $B(S_{3}^{\pm 4/3} \rightarrow j e)=1$, $B(S_{3}^{\pm 2/3} \rightarrow j \nu)=1$, $B(S_{3}^{\pm 1/3} \rightarrow j e)=1/2$, and $B(S_{3}^{\pm 1/3} \rightarrow j \nu)=1/2$.

If we are to use the ATLAS Collaboration results on the $pp \rightarrow \mathrm{LQ} \overline{\mathrm{LQ}}\rightarrow jjee$ process~\cite{ATLAS:2020dsk} to generate accurate constraints on the $S_3$ parameter space, we need to take into account several factors. Namely, in the regime of the QCD dominated leptoquark pair production, i.e., for small $y^{LL}_{3\,11}$, there are two different processes that yield the $jjee$ final state. These are $pp \rightarrow S_{3}^{+4/3} S_{3}^{-4/3} \rightarrow jj ee$ and $pp \rightarrow S_{3}^{+1/3} S_{3}^{-1/3} \rightarrow jj ee$, where the $S_{3}^{+4/3} S_{3}^{-4/3}$ pair goes exclusively into $jj ee$ whereas the $S_{3}^{+1/3} S_{3}^{-1/3}$ pair decays into $jj ee$ only 25\% of the time. If $y^{LL}_{3\,11}$ is not small, we need to include conventional $t$-channel contributions that were discussed in the context of Eq.~\eqref{eq:cross}. These contributions, however, are not the same for $pp \rightarrow S_{3}^{+4/3} S_{3}^{-4/3}$ and $pp \rightarrow S_{3}^{+1/3} S_{3}^{-1/3}$ since the former process is $d\overline{d}$ initiated whereas the latter one is both $u\overline{u}$ and $d\overline{d}$ initiated. Moreover, the $S_{3}^{\pm 4/3}$ couplings to the quark-lepton pairs are always a factor of $\sqrt{2}$ larger than that of $S_{3}^{\pm 1/3}$ due to the $SU(2)$ symmetry of the SM. If we account for all these intricacies, we obtain the Yukawa dependent limit given in Fig.~\ref{fig:LIMITS2} with a thick dashed red curve. A vertical thin dashed red line in Fig.~\ref{fig:LIMITS2}, on the other hand, denotes a naive limit if we were to use purely QCD dominated leptoquark pair production cross sections and yields $m_\mathrm{LQ} \geq 1830$\,GeV. We opted not to present the $S_3$ results with all other scenarios discussed previously in Fig.~\ref{fig:LIMITS1} in order to provide ease of readability. Note, however, that we provide in Fig.~\ref{fig:LIMITS2} limits on the $S_1$ scenarios that were already discussed in Sec.~\ref{sec:S1} and also presented in Fig.~\ref{fig:LIMITS1} for comparison purposes. 

Our considerations, up to this point, did not incorporate potential asymmetric production contributions towards the $jjee$ final state. There are, in general, four asymmetric pair production contributions in any $S_3$ scenario and we present associated diagrams in Fig.~\ref{fig:S3}. Two diagrams in the second row of Fig.~\ref{fig:S3} can give the $jjee$ final state via $pp \rightarrow S_{3}^{-4/3} S_{3}^{+1/3} \rightarrow jjee$ and $pp \rightarrow S_{3}^{+4/3} S_{3}^{-1/3} \rightarrow jjee$ with 50\% probability, each, where $pp \rightarrow S_{3}^{-4/3} S_{3}^{+1/3}$ and $pp \rightarrow S_{3}^{+4/3} S_{3}^{-1/3}$ are $u\overline{d}$ and $d\overline{u}$ initiated, respectively. If we account for these effects, we obtain a limit given in Fig.~\ref{fig:LIMITS2} with a thick dot-dashed red curve. It is this limit that represents correct interpretation of the ATLAS Collaboration results on the $pp \rightarrow \mathrm{LQ} \overline{\mathrm{LQ}}\rightarrow jjee$ process~\cite{ATLAS:2020dsk} when $y^{LL}_{3\,11} \equiv y$ of Eq.~\eqref{eq:main_S_3} is the only non-zero Yukawa coupling. Again, the importance of the asymmetric production inclusion is self-evident. 

\begin{figure}[th!]\centering
\includegraphics[width=1\textwidth]{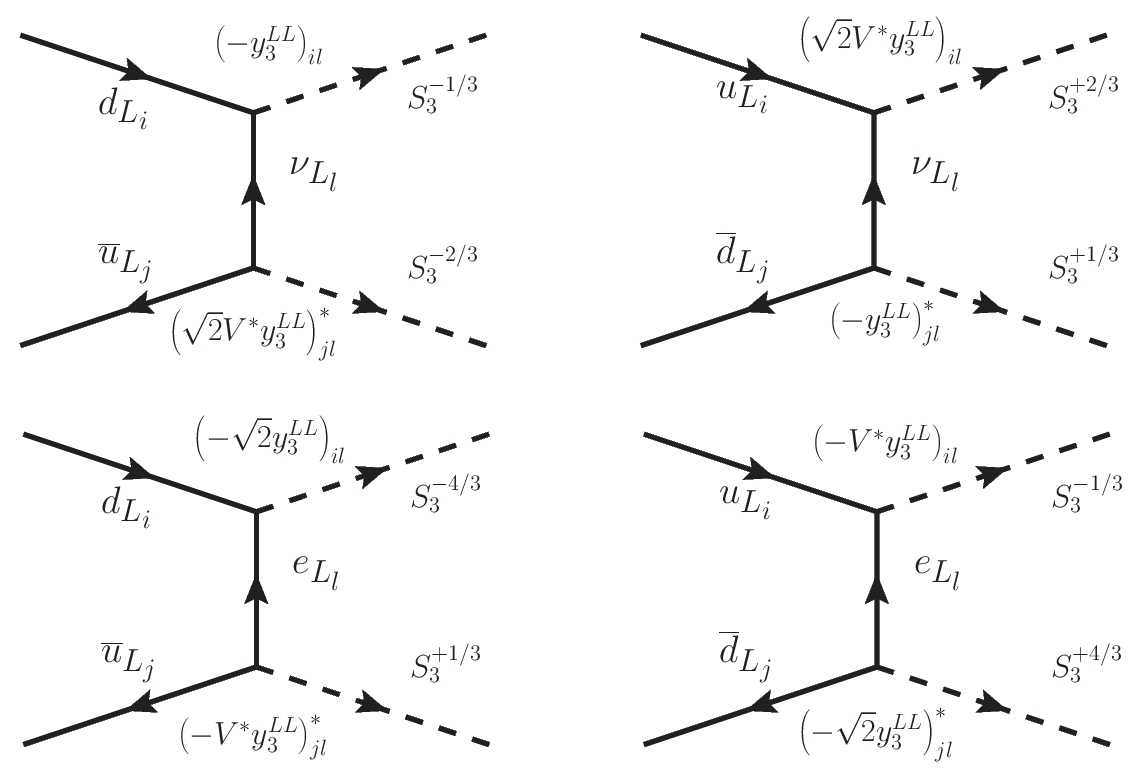}
\caption{Asymmetric pair production for the case of $S_3$ leptoquark.}\label{fig:S3}
\end{figure} 

\item We assume that $y^{LL}_{3\,12} \equiv y$ of Eq.~\eqref{eq:main_S_3} is the only non-zero Yukawa coupling and take all three leptoquarks within $S_3$ multiplet to be degenerate. The branching fractions for the $S_3$ components, when $y^{LL}_{3\,12} \neq 0$, are $B(S_{3}^{\pm 4/3} \rightarrow j \mu)=1$, $B(S_{3}^{\pm 2/3} \rightarrow j \nu)=1$, $B(S_{3}^{\pm 1/3} \rightarrow j \mu)=1/2$, and $B(S_{3}^{\pm 1/3} \rightarrow j \nu)=1/2$. Following the procedure already outlined for the recast of the $y^{LL}_{3\,11} \equiv y \neq 0$ case and applying it on the results of the ATLAS Collaboration search for the $pp \rightarrow \mathrm{LQ} \overline{\mathrm{LQ}}\rightarrow jj\mu\mu$ process~\cite{ATLAS:2020dsk}, we obtain limits rendered in red in Fig.~\ref{fig:LIMITS4}. These converge at $m_\mathrm{LQ} \geq 1770$\,GeV for small values of $y^{LL}_{3\,12} \equiv y$.
\end{itemize}

\subsubsection{Case study: $S_1(\overline{\mathbf{3}},\mathbf{1},1/3)$+$S_3(\overline{\mathbf{3}},\mathbf{3},1/3)$}
\label{sec:S1S3}

Since $S_3$ couples exclusively to the left-chiral leptons we will assume that the only non-zero Yukawa couplings in this scenario, comprising $S_1$ and $S_3$ leptoquarks, are either $y^{LL}_{1\,11}$ and $y^{LL}_{3\,11}$ or $y^{LL}_{1\,12}$ and $y^{LL}_{3\,12}$. This ansatz will allow us to present an analysis of the asymmetric pair production effects within the $\Delta F=0$ system.

\begin{itemize}
\item  We first consider scenario with $y^{LL}_{1\,11} \neq 0$ and $y^{LL}_{3\,11} \neq 0$. The branching fractions of leptoquarks are $B(S_{3}^{\pm 4/3} \rightarrow j e)=1$, $B(S_{3}^{\pm 2/3} \rightarrow j \nu)=1$, $B(S_{3}^{\pm 1/3} \rightarrow j e)=1/2$, $B(S_{3}^{\pm 1/3} \rightarrow j \nu)=1/2$, $B(S_{1}^{\pm 1/3} \rightarrow j e)=1/2$, and $B(S_{1}^{\pm 1/3} \rightarrow j \nu)=1/2$. We furthermore assume that $S_1$ and the components of $S_3$ are degenerate in mass and also take that $y^{LL}_{1\,11} = y^{LL}_{3\,11} \equiv y$ to simplify discussion.

A naive QCD limit on the parameter space of this scenario, set by the ATLAS Collaboration data on the $pp \rightarrow \mathrm{LQ} \overline{\mathrm{LQ}}\rightarrow jjee$ process~\cite{ATLAS:2020dsk}, is presented in Fig.~\ref{fig:LIMITS2} with a vertical thin dashed green line and corresponds to $m_\mathrm{LQ} \geq 1860$\,GeV. If we also include the usual $t$-channel contributions for both $S_1$ and $S_3$, as discussed previously in Secs.~\ref{sec:S1} and \ref{sec:S3}, respectively, we obtain the limit given with a thick dashed green curve in Fig.~\ref{fig:LIMITS2}.

In order to numerically evaluate the asymmetric pair production contributions we need to account for $pp \rightarrow S_{3}^{-4/3} S_{3}^{+1/3} \rightarrow jjee$ (50\%), $pp \rightarrow S_{3}^{+4/3} S_{3}^{-1/3} \rightarrow jjee$ (50\%), $pp \rightarrow S_{3}^{-4/3} S_{1}^{+1/3} \rightarrow jjee$ (50\%), $pp \rightarrow S_{3}^{+4/3} S_{1}^{-1/3} \rightarrow jjee$ (50\%), $pp \rightarrow S_{3}^{-1/3} S_{1}^{+1/3} \rightarrow jjee$ (25\%) and $pp \rightarrow S_{3}^{+1/3} S_{1}^{-1/3} \rightarrow jjee$ (25\%), where we specify in parentheses the associated decay rate into the $jjee$ final state for each of these processes. Note that the last two processes are both $u\overline{u}$ and $d\overline{d}$ initiated. In fact, $S_1$+$S_3$ scenario is the only $\Delta F=0$ scenario that features asymmetric production initiated with the $q\overline{q^\prime}$ combination, where both $q$ and $q^\prime$ are of the same type of flavor. Moreover, these  same-flavor contributions always come in pairs as they are simultaneously generated by the up-type and down-type quarks. This is the reason why we opted to present combinations $\sigma^\mathrm{pair}_{u\overline{u}}+\sigma^\mathrm{pair}_{d\overline{d}}$, $\sigma^\mathrm{pair}_{u\overline{c}}+\sigma^\mathrm{pair}_{d\overline{s}}$, $\sigma^\mathrm{pair}_{c\overline{u}}+\sigma^\mathrm{pair}_{s\overline{d}}$ and $\sigma^\mathrm{pair}_{c\overline{c}}+\sigma^\mathrm{pair}_{s\overline{s}}$ in Fig.~\ref{fig:C_S_qaq} instead of individual $q\overline{q^\prime}$ contributions. 

If we properly include all the relevant processes that yield the $jjee$ final state, we obtain a limit on the $S_1$+$S_3$ scenario parameter space that is given by a thick dot-dashed green curve in Fig.~\ref{fig:LIMITS2}. The parameter space to the left of that curve is excluded by the ATLAS Collaboration search for the $pp \rightarrow \mathrm{LQ} \overline{\mathrm{LQ}} \rightarrow jjee$ process~\cite{ATLAS:2020dsk}.

\item If we assume that $y^{LL}_{1\,12} \neq 0$ and $y^{LL}_{3\,12} \neq 0$, the branching fractions of leptoquarks read $B(S_{3}^{\pm 4/3} \rightarrow j \mu)=1$, $B(S_{3}^{\pm 2/3} \rightarrow j \nu)=1$, $B(S_{3}^{\pm 1/3} \rightarrow j \mu)=1/2$, $B(S_{3}^{\pm 1/3} \rightarrow j \nu)=1/2$, $B(S_{1}^{\pm 1/3} \rightarrow j \mu)=1/2$, and $B(S_{1}^{\pm 1/3} \rightarrow j \nu)=1/2$. If we also assume that $S_1$ and the components of $S_3$ are degenerate in mass and take that $y^{LL}_{1\,12} = y^{LL}_{3\,12} \equiv y$, we obtain the set of limits rendered in green in Fig.~\ref{fig:LIMITS4} that converge to $m_\mathrm{LQ} \geq 1800$\,GeV for small values of $y^{LL}_{1\,12} = y^{LL}_{3\,12} \equiv y$. These limits use the ATLAS Collaboration search results for the $pp \rightarrow \mathrm{LQ} \overline{\mathrm{LQ}} \rightarrow jj\mu\mu$ process~\cite{ATLAS:2020dsk}. A vertical thin dashed line is the bound based on the QCD cross section. A thick dashed curve is generated if one accounts for the usual $t$-channel contributions whereas a thick dot-dashed curve is the accurate limit that incorporates both the conventional and asymmetric leptoquark pair production mechanism effects.
\end{itemize}

\subsection{Final remarks}

Before we conclude this section, several remarks are in order. 
\begin{itemize}
    \item We have explicitly assumed in our analysis that the asymmetrically produced leptoquarks are mass degenerate. If that is not the case the asymmetric pair production mechanism would allow for an unambiguous and unique test of existence of multiple leptoquarks if the leptoquarks in question couple to the lepton(s) of the same chirality. The asymmetric pair production search would thus be complementary to other detection methods, either direct or indirect, to ascertain the existence of these hypothetical particles at hadron colliders.  
    \item We have not included the CKM matrix effects in our numerical study. We have, however, commented in Sec.~\ref{sec:S1} on the fact that the CKM matrix effects can reduce the branching ratio of specific channels we considered, thereby affecting the associated bounds on the leptoquark parameter space. We have also not performed a full next-to-leading order simulation of the leptoquark pair production cross sections. These effects can be accounted for with available tools but would only affect quantitative aspect of our study without compromising our main message. 
    \item It is important to note that all leptoquark scenarios that we presented require individual attention if one is to extract accurate parameter space constraints. In fact, even in the small Yukawa coupling limit, different scenarios would usually yield different lower bounds on the mass of relevant leptoquark(s). Our study should thus be seen as a blueprint for inclusion of the asymmetric leptoquark pair production effects and proper interpretation of available experimental data. 
\end{itemize}

\section{Atomic parity violation}
\label{sec:APV}

In the proceeding section we produce leptoquark pair production search limits for various scalar leptoquark scenarios when the leptoquarks in question exclusively couple to either electrons or muons and the first generation quarks. In the former case it is also important to address the impact of the APV search constraints on otherwise viable leptoquark parameter space. 

The effective APV interactions can be parametrized as~\cite{Langacker:1990jf}
\begin{align}
\label{eq:APV_L}
\mathcal{L}_\mathrm{PV}=\frac{G_F}{\sqrt{2}} \left(\overline e \gamma^\mu \gamma^5 e\right) \left(\sum_{q=u,d} \hat C_{1q} \overline q \gamma_\mu q
\right),     
\end{align}
where coefficients $\hat C_{1q}=C^\mathrm{SM}_{1q}+C^\mathrm{NP}_{1q}$ capture both the the SM and the New Physics (NP) contributions. In particular, $C^\mathrm{SM}_{1u}=-0.1887$ and $C^\mathrm{SM}_{1d}=0.3419$~\cite{Qweak:2018tjf}, whereas the NP contributions $C^\mathrm{NP}_{1q}$ for our scenarios are given in Table~\ref{tab:700}~\cite{Barger:2000gv,Crivellin:2021egp}. 

\begin{table}[th!]
\begin{center}
\begin{tabular}{|c|c|c|c|}\hline
Non-zero Yukawas & $C_{1u}^\mathrm{NP}$ & $C_{1d}^\mathrm{NP}$ & LQ scenario \\\hline\hline
 
$y_{1\;11}^\mathrm{LL} \equiv y$ & $-\frac{v^2}{4 m^2_\mathrm{LQ}}  \left|y\right|^2$ & $0$ & \multirow{2}{1cm}{~~$S_1$}\\ \cline{1-3}
$y_{1\;11}^\mathrm{RR} \equiv y$ & $\frac{v^2}{4 m^2_\mathrm{LQ}} \left|y\right|^2$ & $0$ & \\ \hline\hline

$y_{2\;11}^\mathrm{LR} \equiv y$ & $-\frac{v^2}{4 m^2_\mathrm{LQ}}  \left|y\right|^2$ & $-\frac{v^2}{4 m^2_\mathrm{LQ}}  \left|y\right|^2$ & \multirow{2}{1cm}{~~$R_2$}\\ \cline{1-3}
$y_{2\;11}^\mathrm{RL} \equiv y$ & $\frac{v^2}{4 m^2_\mathrm{LQ}} \left|y\right|^2$ & $0$ & \\ \hline\hline

$y_{1\;11}^\mathrm{RR}=y_{2\;11}^\mathrm{LR} \equiv y$ & $0$ & $-\frac{v^2}{4 m^2_\mathrm{LQ}} \left|y\right|^2$ & $S_1$+$R_2$ \\ \hline\hline

$y_{3\;11}^\mathrm{LL} \equiv y$ & $-\frac{v^2}{4 m^2_\mathrm{LQ}} \left|y\right|^2$ & $-\frac{v^2}{2 m^2_\mathrm{LQ}} \left|y\right|^2$ & $S_3$ \\ \hline\hline

$y_{1\;11}^\mathrm{LL}=y_{3\;11}^\mathrm{LL} \equiv y$ & $-\frac{v^2}{2 m^2_\mathrm{LQ}} \left|y\right|^2$ & $-\frac{v^2}{2 m^2_\mathrm{LQ}} \left|y\right|^2$ & $S_1$+$S_3$ \\
 \hline
\end{tabular}
\end{center}
\caption{$C^\mathrm{NP}_{1q}$ coefficients~\cite{Barger:2000gv,Crivellin:2021egp} for the scalar leptoquark scenarios under consideration. The mass and Yukawa coupling degeneracies of leptoquarks are understood while $v=246$\,GeV.}
\label{tab:700}
\end{table}

\begin{figure}[t!]\centering
\includegraphics[width=0.48\textwidth]{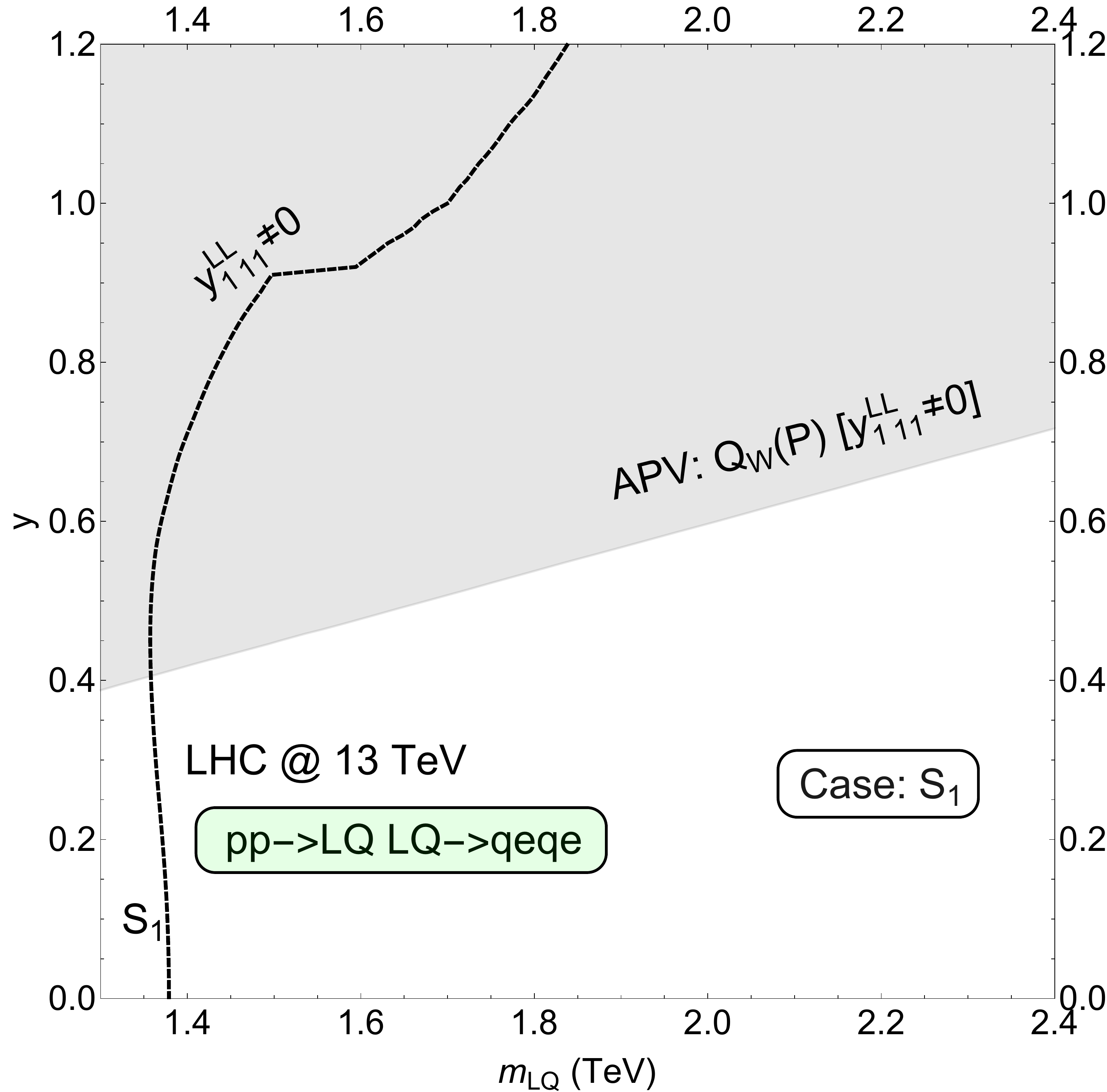}
\includegraphics[width=0.48\textwidth]{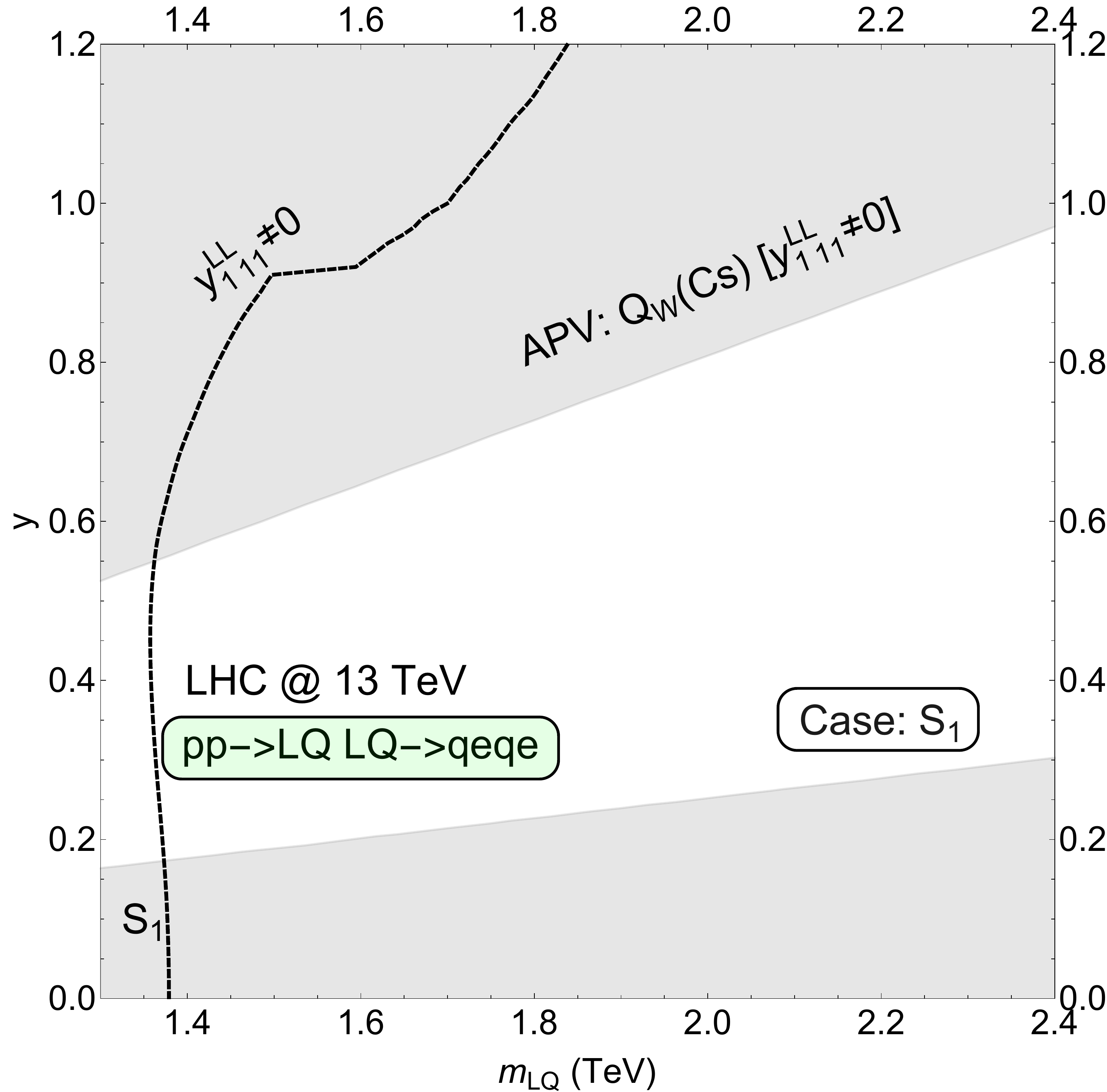}
\\ \vspace{10pt}
\includegraphics[width=0.48\textwidth]{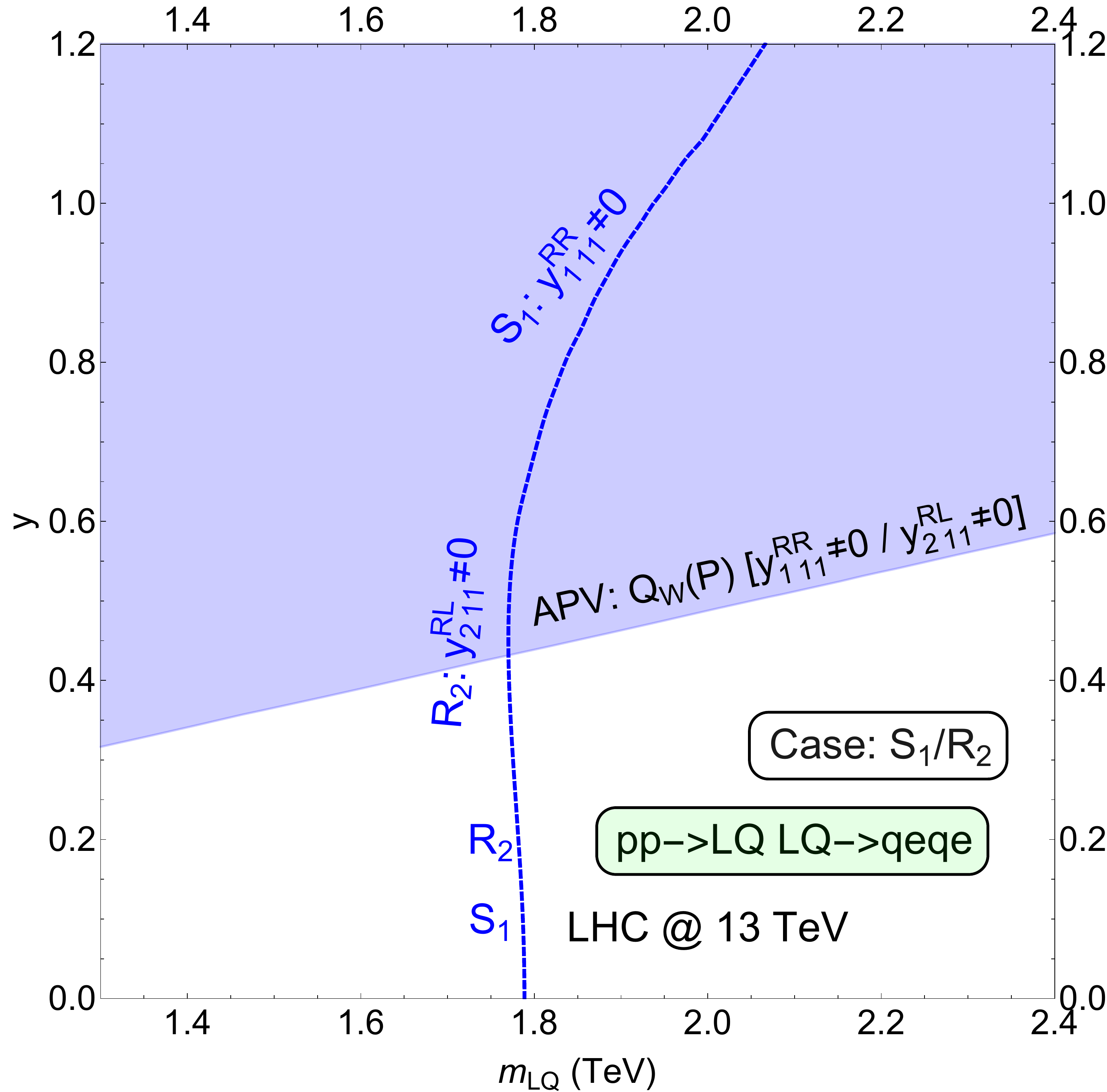}
\includegraphics[width=0.48\textwidth]{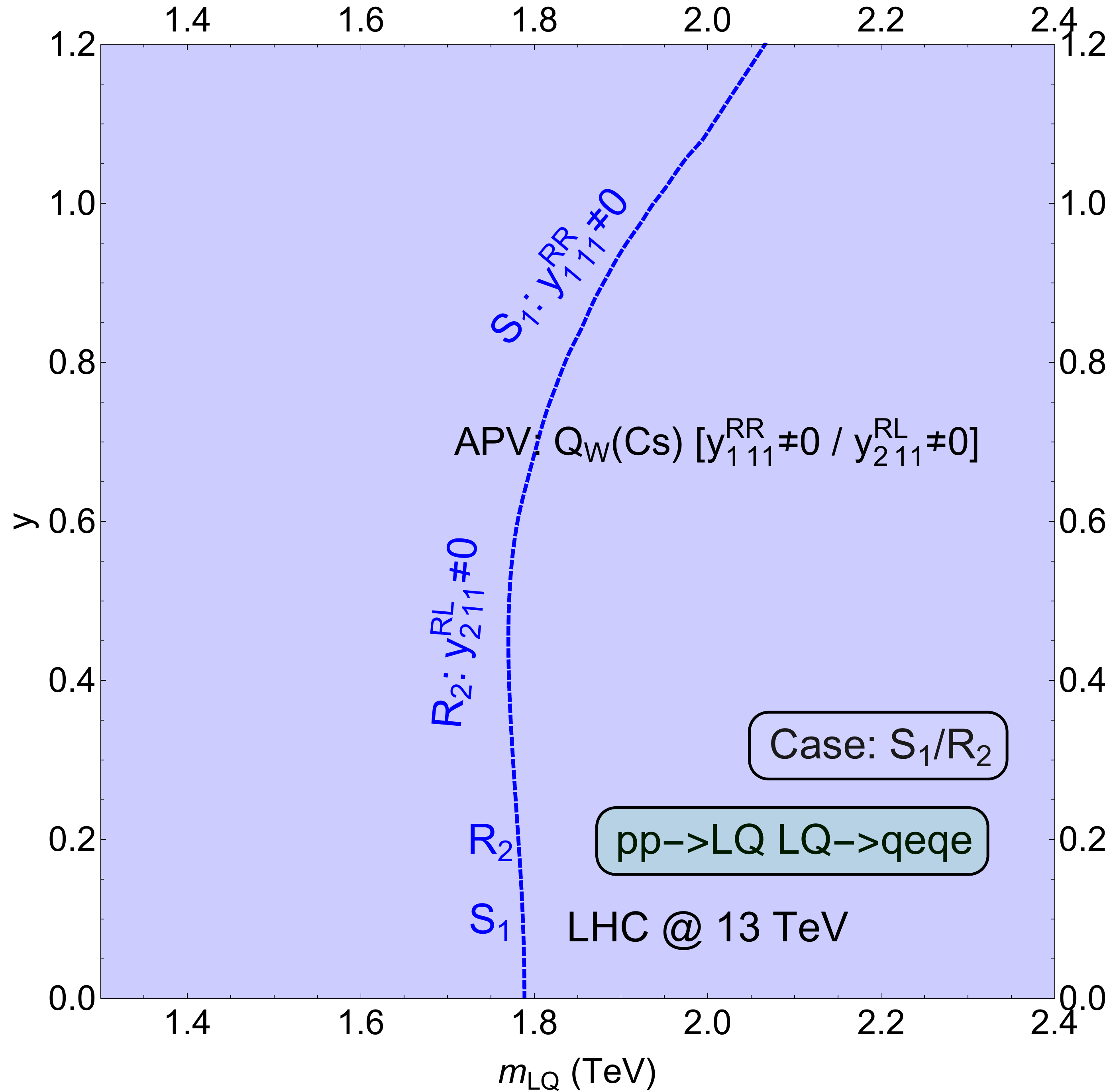}
\caption{APV limits for the proton (left) and Cs (right) measurements juxtaposed with the leptoquark pair production search limits for the corresponding leptoquark scenarios, as indicated. Shaded regions are ruled out by the APV measurements.}\label{fig:APV01}
\end{figure} 

The content of Table~\ref{tab:700} clearly shows that each leptoquark scenario, except for the $S_1$ scenario with $y_{1\;11}^\mathrm{RR} \equiv y$ and $R_2$ with $y_{2\;11}^\mathrm{RL} \equiv y$ that are identical, is to be treated differently when it comes to APV constraints. This situation exactly mirrors our findings with regard to constraints originating from the leptoquark pair production search. We accordingly have six distinct cases to consider, all in all.

\begin{figure}[th!]\centering
\includegraphics[width=0.48\textwidth]{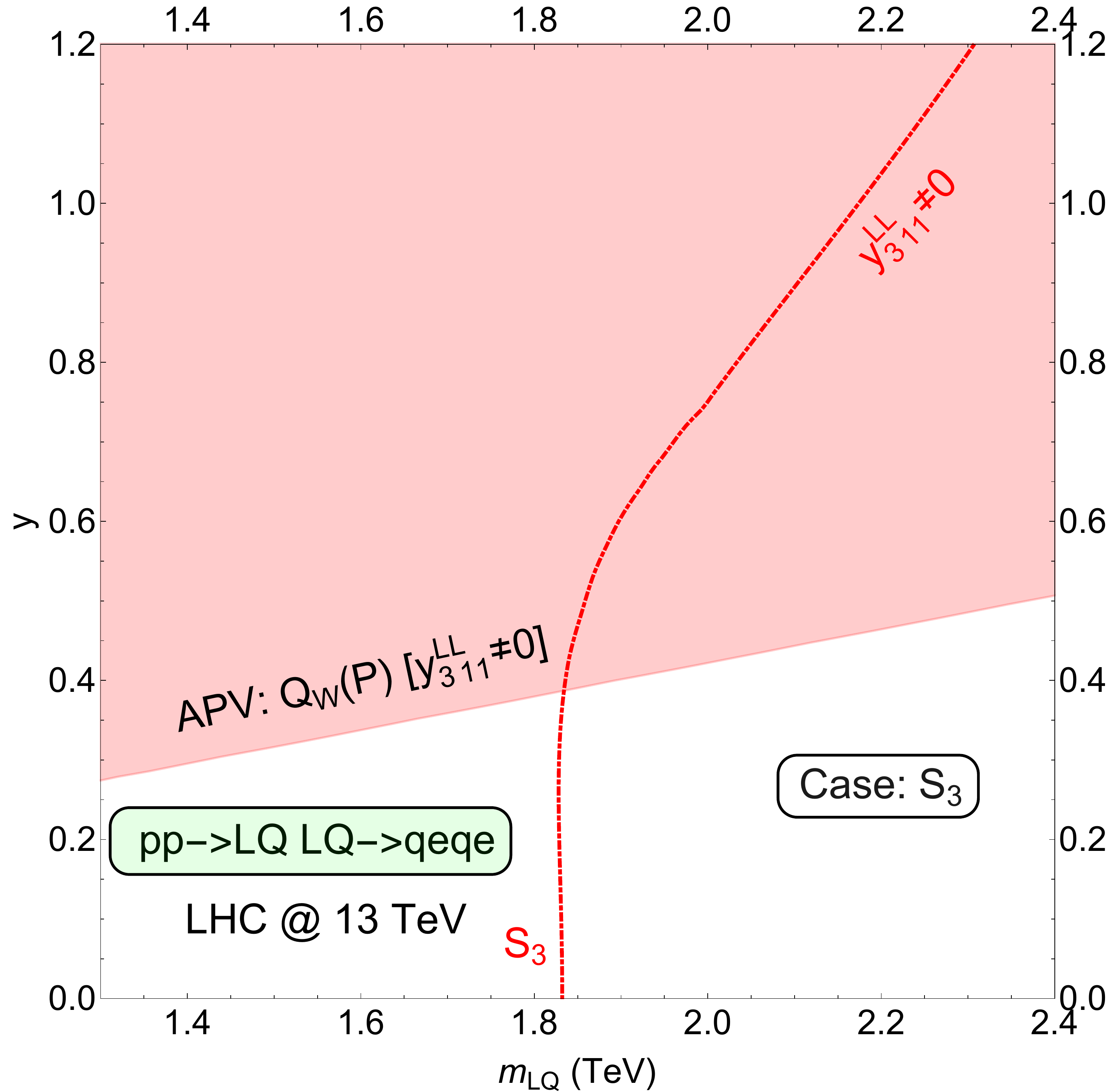}
\includegraphics[width=0.48\textwidth]{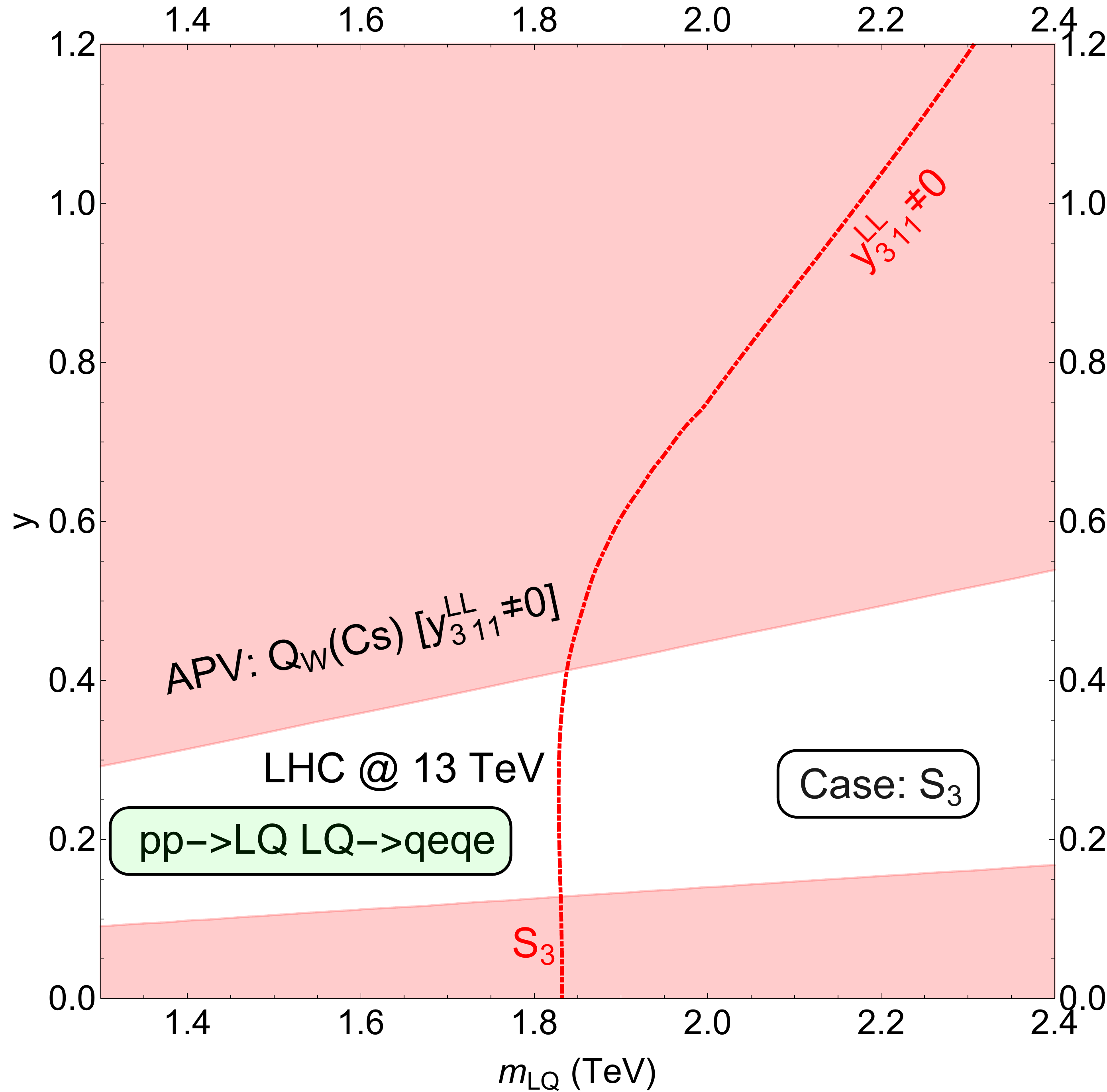}
\\ \vspace{10pt}
\includegraphics[width=0.48\textwidth]{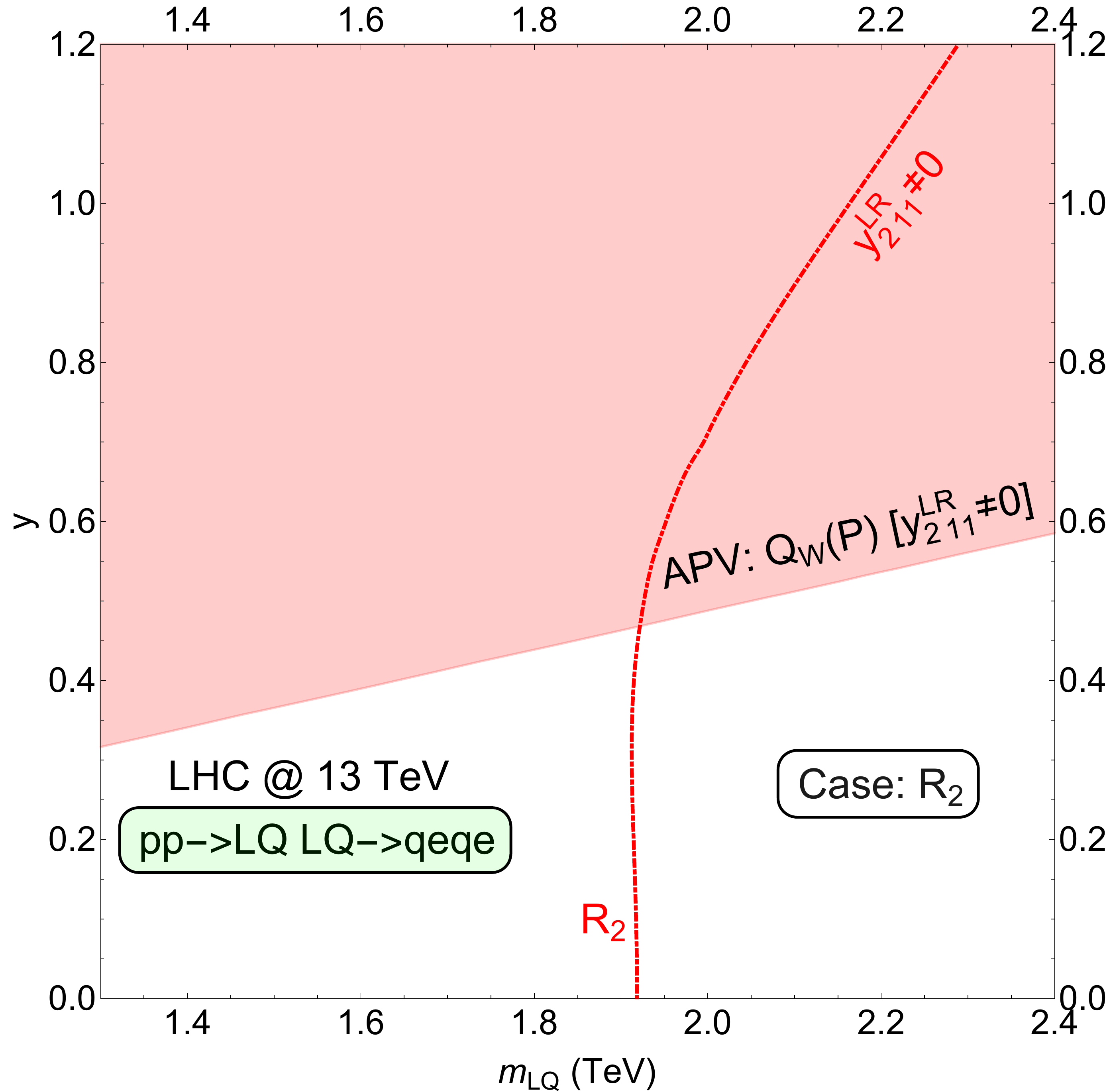}
\includegraphics[width=0.48\textwidth]{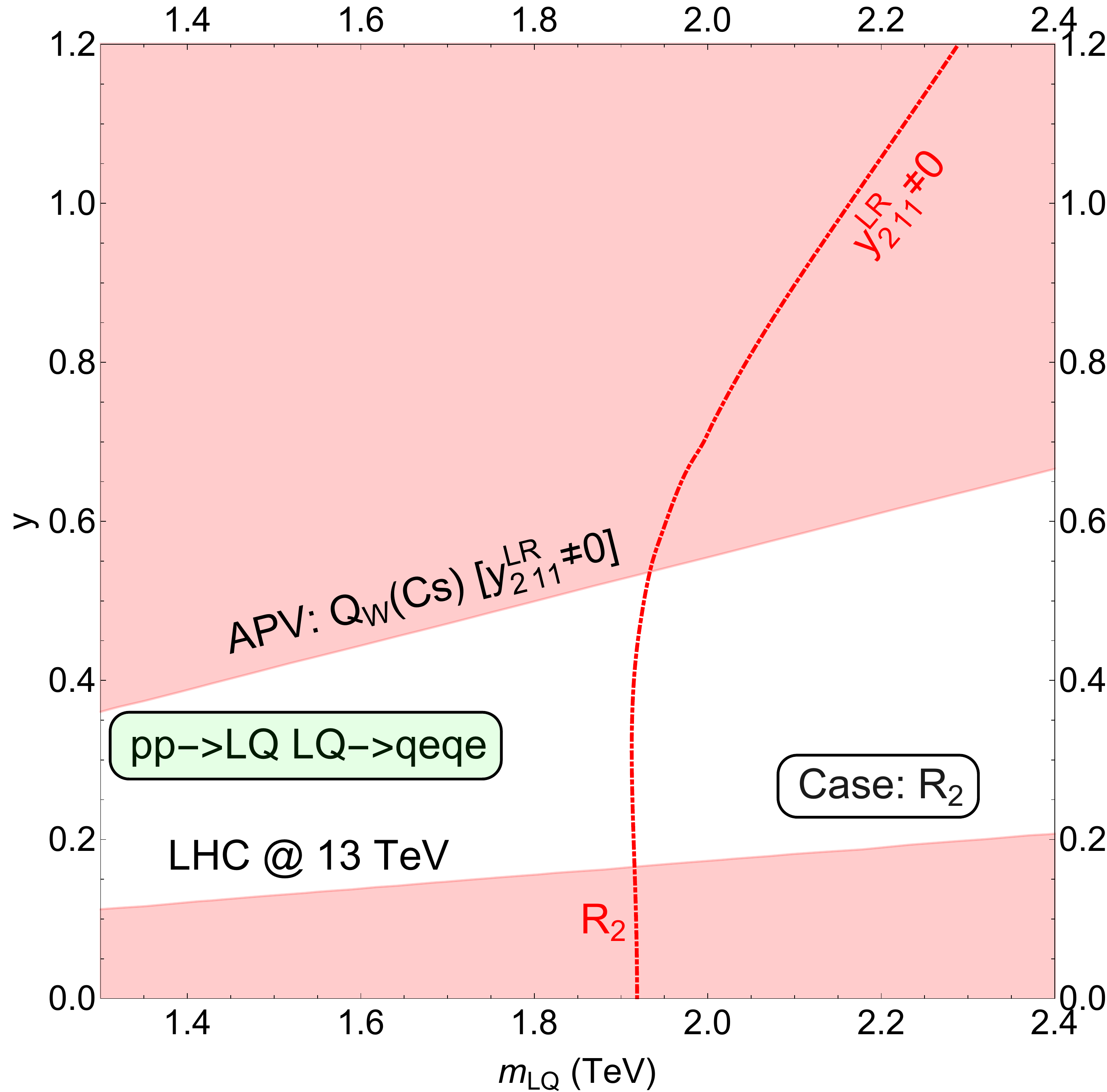}
\caption{APV limits for the proton (left) and Cs (right) measurements juxtaposed with the leptoquark pair production search limits for the corresponding leptoquark scenarios, as indicated. Shaded regions are ruled out by the APV measurements.}\label{fig:APV02}
\end{figure} 

\begin{figure}[th!]\centering
\includegraphics[width=0.48\textwidth]{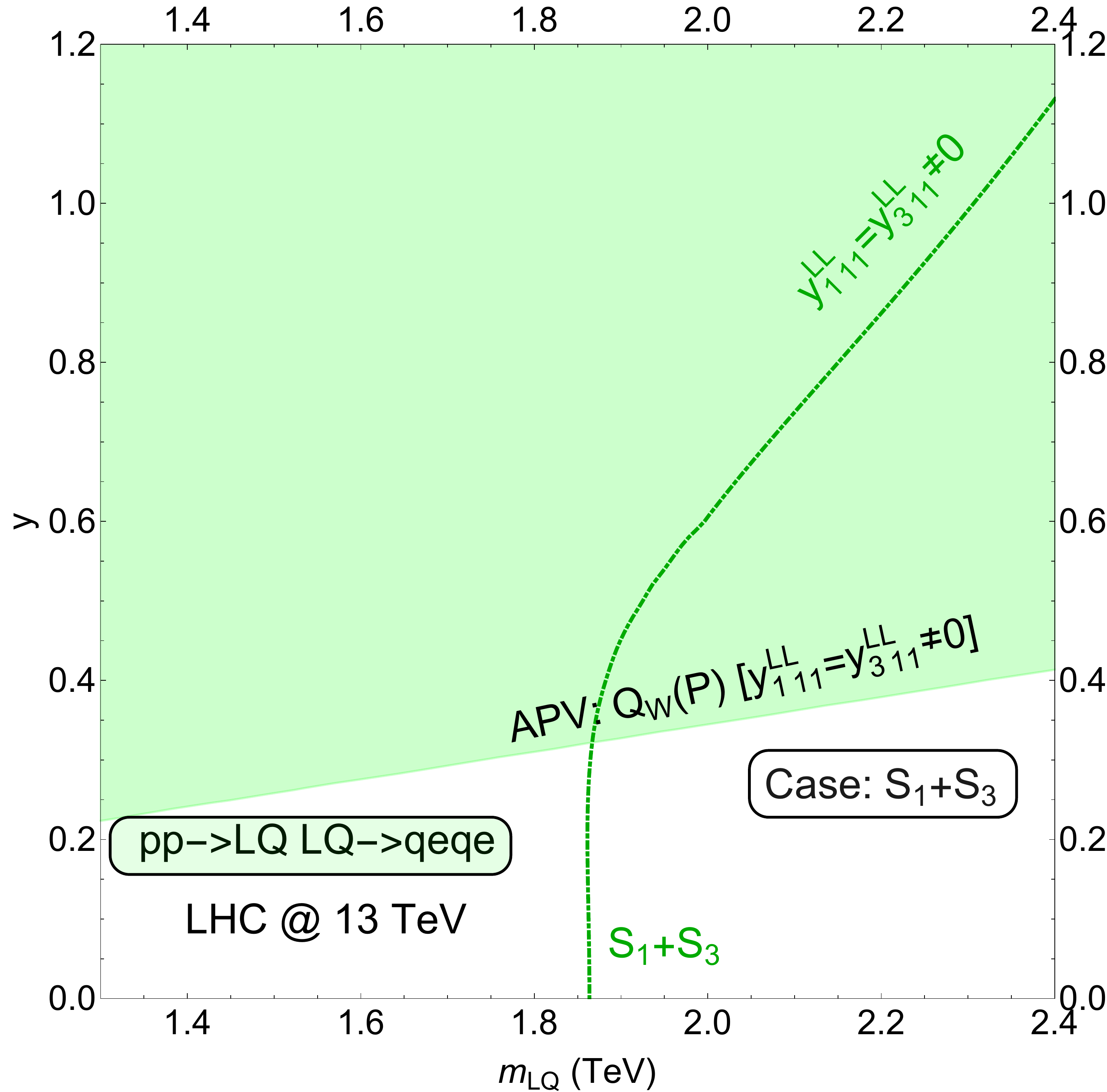}
\includegraphics[width=0.48\textwidth]{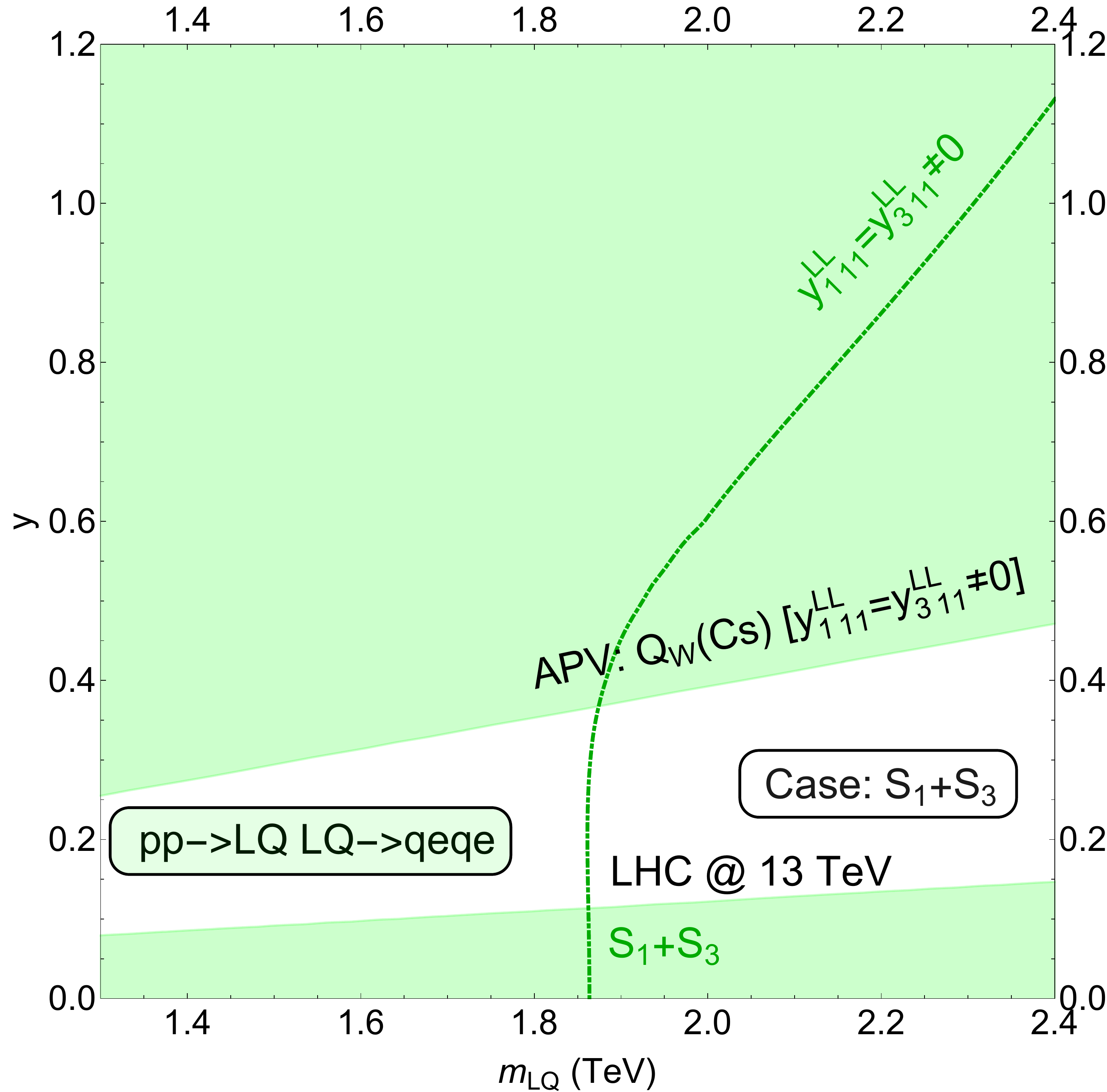}
\\ \vspace{10pt}
\includegraphics[width=0.48\textwidth]{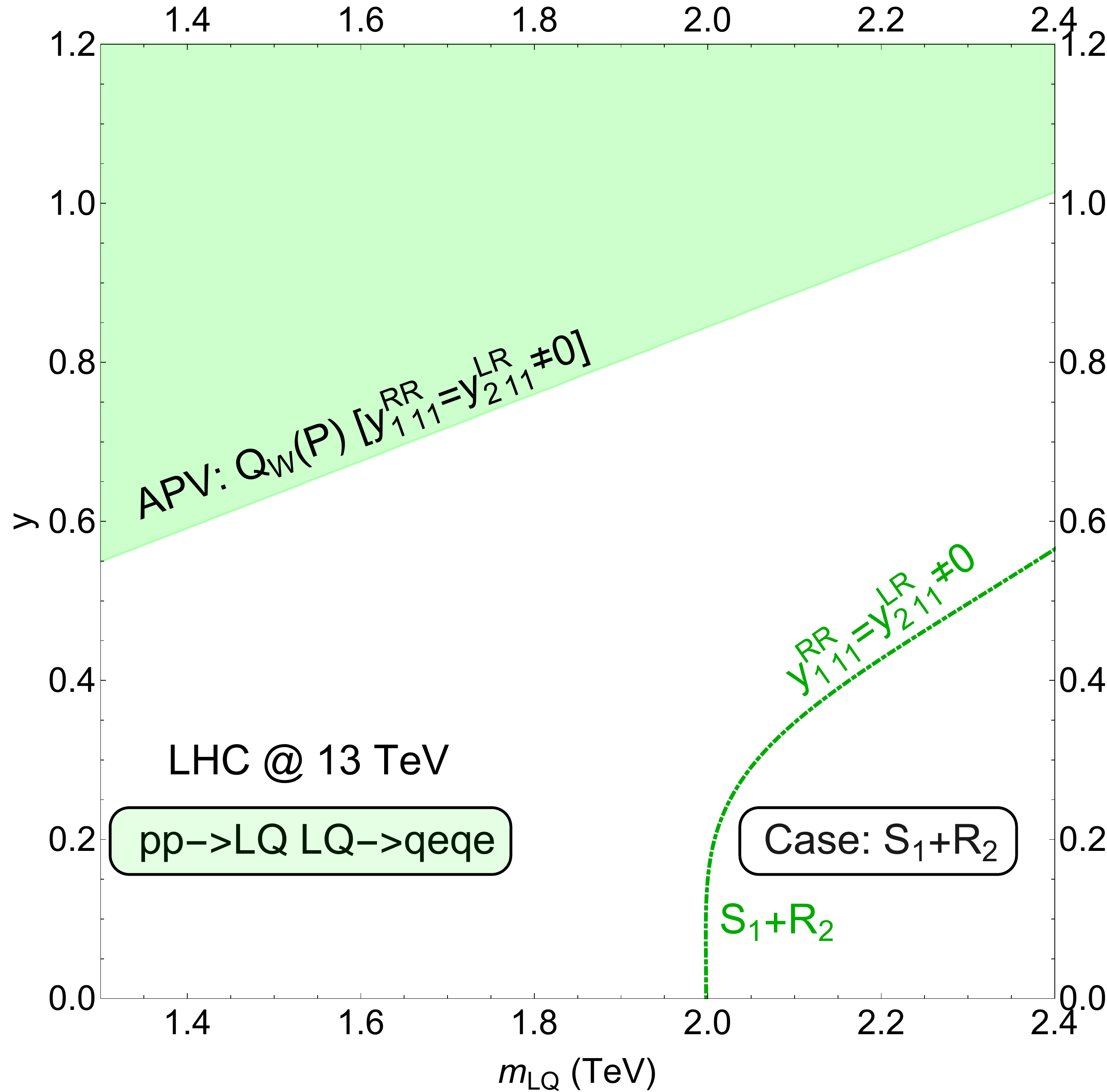}
\includegraphics[width=0.48\textwidth]{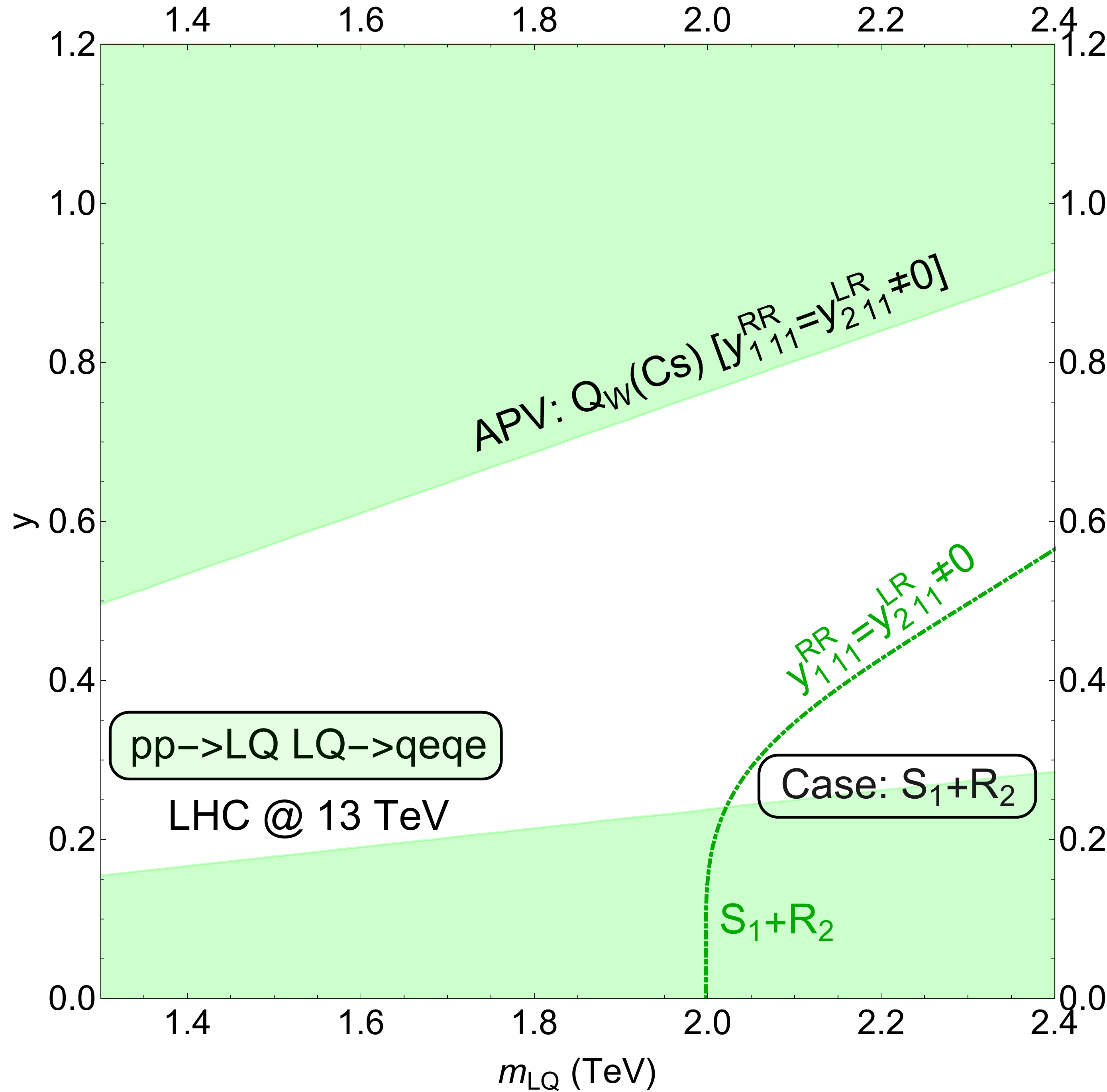}
\caption{APV limits for the proton (left) and Cs (right) measurements juxtaposed with the leptoquark pair production search limits for the corresponding leptoquark scenarios, as indicated. Shaded regions are ruled out by the APV measurements.}\label{fig:APV03}
\end{figure}

To proceed one defines a nuclear weak charge  
\begin{align}
Q_W\left(Z,N\right)=-2\left(2Z+N\right)2 \hat C_{1u}  -2\left(Z+2 N\right) \hat C_{1d},   
\end{align}
where $Z$ is a nuclear charge number and $N$ represents a number of neutrons. 
The experimental measurements of the nuclear weak charge of the proton ($Q_\mathrm{W}(p)$) and ${}^{133}\mathrm{Cs}$ ($Q_\mathrm{W}({}^{133}\mathrm{Cs})$) are~\cite{ParticleDataGroup:2020ssz,Crivellin:2021egp}
\begin{align}
Q_\mathrm{W}(p)=-2\left( 2\hat C_{1u}+\hat C_{1d} \right)= 0.0719\pm 0.0045,    \end{align}
and 
\begin{align}
Q_\mathrm{W}({}^{133}\mathrm{Cs})=-2\left( 188\hat C_{1u}+211 \hat C_{1d} \right)= -72.82\pm 0.42,    
\end{align}
respectively. It is important to note that the measurement of $Q_\mathrm{W}(p)$ is in agreement with the SM prediction whereas the measured value of $Q_\mathrm{W}({}^{133}\mathrm{Cs})$ is not. In fact, the value of $Q_\mathrm{W}({}^{133}\mathrm{Cs})$ prefers finite negative NP contributions, as, for example, given with negative entries in Table~\ref{tab:700}. We accordingly opt to show separately constraints generated by measurements of $Q_\mathrm{W}(p)$ and $Q_\mathrm{W}({}^{133}\mathrm{Cs})$ in Figs.~\ref{fig:APV01}, \ref{fig:APV02}, and \ref{fig:APV03}, juxtaposing them with the leptoquark pair production search limits, where we also group aforementioned six different leptoquark scenarios pairwise. The first column in Figs.~\ref{fig:APV01}, \ref{fig:APV02}, and \ref{fig:APV03} is reserved for the $Q_\mathrm{W}(p)$ generated constraint, whereas the second column reflect the impact of the $Q_\mathrm{W}({}^{133}\mathrm{Cs})$ measurement on the leptoquark parameter space. The shaded regions in Figs.~\ref{fig:APV01}, \ref{fig:APV02}, and \ref{fig:APV03} are ruled out by the APV measurements at the $1\sigma$ level, where different leptoquark scenarios are shown in separate rows for clarity. 

Fig.~\ref{fig:APV03} clearly shows that the pair production constraint for the $S_1$+$R_2$ scenario is superior to the existing APV constraints. It is also clear that the the $S_1$ scenario with $y_{1\;11}^\mathrm{RR} \neq 0$ and $R_2$ with $y_{2\;11}^\mathrm{RL} \neq 0$ are completely ruled out by the $Q_\mathrm{W}({}^{133}\mathrm{Cs})$ measurement. Of course, the APV constraints are irrelevant for the scenarios when leptoquarks couple to muons. 
We also note that it is possible to arrange for cancellation between individual leptoquark contributions towards APV interactions of Eq.~\eqref{eq:APV_L}. This possibility of having the NP coefficients $C^\mathrm{NP}_{1q}$ vanish can be, for example, trivially realised within the $S_1$ scenario with $y_{1\;11}^\mathrm{LL} = y_{1\;11}^\mathrm{RR} \equiv y$ or within the $S_1$+$R_2$ scenario with $y_{1\;11}^\mathrm{LL} = y_{2\;11}^\mathrm{RL} \equiv y$.

\section{Conclusions}
\label{sec:CONCLUSIONS}

This work investigates the asymmetric leptoquark pair production mechanism at the LHC. A sharp difference between the conventional leptoquark pair production and the asymmetric one is that for the latter, which is produced via $t$-channel lepton exchange, the pairs of produced leptoquarks are not conjugate states of each other. We spell out necessary conditions for an operational asymmetric leptoquark pair production mechanism and catalog all possible combinations of leptoquark multiplets that can potentially generate it. We, furthermore,  demonstrate how to properly combine asymmetric and conventional pair production mechanism effects by considering several scenarios where the SM is extended with either one or two scalar leptoquark multiplets. Finally, based on our analysis, we advocate that contributions from asymmetric pair production should be included when deriving reliable constraints on leptoquark parameter space as well as be used when attempting to perform correct identification of these promising new physics sources. 

\acknowledgments

I.D.\ would like to thank Svjetlana Fajfer for numerous fruitful discussions with regard to this project. 

\bibliographystyle{style}
\bibliography{ref}

\providecommand{\href}[2]{#2}\begingroup\raggedright\begin{thebibliography}{10}

\bibitem{ATLAS:2020dsk}
{\bfseries ATLAS} Collaboration, G.~Aad {\em et~al.}, ``{Search for pairs of
  scalar leptoquarks decaying into quarks and electrons or muons in $ \sqrt{s}
  $ = 13 TeV $pp$ collisions with the ATLAS detector},''
  \href{http://dx.doi.org/10.1007/JHEP10(2020)112}{{\em JHEP} {\bfseries 10}
  (2020) 112}, \href{http://arxiv.org/abs/2006.05872}{{\ttfamily
  arXiv:2006.05872 [hep-ex]}}.

\bibitem{ATLAS:2020xov}
{\bfseries ATLAS} Collaboration, G.~Aad {\em et~al.}, ``{Search for pair
  production of scalar leptoquarks decaying into first- or second-generation
  leptons and top quarks in proton\textendash{}proton collisions at $\sqrt{s}$
  = 13 TeV with the ATLAS detector},''
  \href{http://dx.doi.org/10.1140/epjc/s10052-021-09009-8}{{\em Eur. Phys. J.
  C} {\bfseries 81} no.~4, (2021) 313},
  \href{http://arxiv.org/abs/2010.02098}{{\ttfamily arXiv:2010.02098
  [hep-ex]}}.

\bibitem{CMS:2020wzx}
{\bfseries CMS} Collaboration, A.~M. Sirunyan {\em et~al.}, ``{Search for
  singly and pair-produced leptoquarks coupling to third-generation fermions in
  proton-proton collisions at s=13~TeV},''
  \href{http://dx.doi.org/10.1016/j.physletb.2021.136446}{{\em Phys. Lett. B}
  {\bfseries 819} (2021) 136446},
  \href{http://arxiv.org/abs/2012.04178}{{\ttfamily arXiv:2012.04178
  [hep-ex]}}.

\bibitem{ATLAS:2021oiz}
{\bfseries ATLAS} Collaboration, G.~Aad {\em et~al.}, ``{Search for pair
  production of third-generation scalar leptoquarks decaying into a top quark
  and a $\tau$-lepton in $pp$ collisions at $ \sqrt{s} $ = 13 TeV with the
  ATLAS detector},'' \href{http://dx.doi.org/10.1007/JHEP06(2021)179}{{\em
  JHEP} {\bfseries 06} (2021) 179},
  \href{http://arxiv.org/abs/2101.11582}{{\ttfamily arXiv:2101.11582
  [hep-ex]}}.

\bibitem{CMS:2022zks}
{\bfseries CMS} Collaboration, ``{The search for a third-generation leptoquark
  coupling to a $\tau$ lepton and a b quark through single, pair and
  nonresonant production at $\sqrt{s}=13~\mathrm{TeV}$},''.

\bibitem{Kramer:1997hh}
M.~Kramer, T.~Plehn, M.~Spira, and P.~M. Zerwas, ``{Pair production of scalar
  leptoquarks at the Tevatron},''
  \href{http://dx.doi.org/10.1103/PhysRevLett.79.341}{{\em Phys. Rev. Lett.}
  {\bfseries 79} (1997) 341--344},
  \href{http://arxiv.org/abs/hep-ph/9704322}{{\ttfamily arXiv:hep-ph/9704322}}.

\bibitem{Kramer:2004df}
M.~Kramer, T.~Plehn, M.~Spira, and P.~M. Zerwas, ``{Pair production of scalar
  leptoquarks at the CERN LHC},''
  \href{http://dx.doi.org/10.1103/PhysRevD.71.057503}{{\em Phys. Rev. D}
  {\bfseries 71} (2005) 057503},
  \href{http://arxiv.org/abs/hep-ph/0411038}{{\ttfamily arXiv:hep-ph/0411038}}.

\bibitem{Mandal:2015lca}
T.~Mandal, S.~Mitra, and S.~Seth, ``{Pair Production of Scalar Leptoquarks at
  the LHC to NLO Parton Shower Accuracy},''
  \href{http://dx.doi.org/10.1103/PhysRevD.93.035018}{{\em Phys. Rev. D}
  {\bfseries 93} no.~3, (2016) 035018},
  \href{http://arxiv.org/abs/1506.07369}{{\ttfamily arXiv:1506.07369
  [hep-ph]}}.

\bibitem{Dorsner:2018ynv}
I.~Dor\v{s}ner and A.~Greljo, ``{Leptoquark toolbox for precision collider
  studies},'' \href{http://dx.doi.org/10.1007/JHEP05(2018)126}{{\em JHEP}
  {\bfseries 05} (2018) 126}, \href{http://arxiv.org/abs/1801.07641}{{\ttfamily
  arXiv:1801.07641 [hep-ph]}}.

\bibitem{Beenakker:2016lwe}
W.~Beenakker, C.~Borschensky, M.~Kr\"amer, A.~Kulesza, and E.~Laenen,
  ``{NNLL-fast: predictions for coloured supersymmetric particle production at
  the LHC with threshold and Coulomb resummation},''
  \href{http://dx.doi.org/10.1007/JHEP12(2016)133}{{\em JHEP} {\bfseries 12}
  (2016) 133}, \href{http://arxiv.org/abs/1607.07741}{{\ttfamily
  arXiv:1607.07741 [hep-ph]}}.

\bibitem{Beenakker:1997ut}
W.~Beenakker, M.~Kramer, T.~Plehn, M.~Spira, and P.~M. Zerwas, ``{Stop
  production at hadron colliders},''
  \href{http://dx.doi.org/10.1016/S0550-3213(98)00014-5}{{\em Nucl. Phys. B}
  {\bfseries 515} (1998) 3--14},
  \href{http://arxiv.org/abs/hep-ph/9710451}{{\ttfamily arXiv:hep-ph/9710451}}.

\bibitem{Beenakker:2010nq}
W.~Beenakker, S.~Brensing, M.~Kramer, A.~Kulesza, E.~Laenen, and I.~Niessen,
  ``{Supersymmetric top and bottom squark production at hadron colliders},''
  \href{http://dx.doi.org/10.1007/JHEP08(2010)098}{{\em JHEP} {\bfseries 08}
  (2010) 098}, \href{http://arxiv.org/abs/1006.4771}{{\ttfamily arXiv:1006.4771
  [hep-ph]}}.

\bibitem{Beenakker:2016gmf}
W.~Beenakker, C.~Borschensky, R.~Heger, M.~Kr\"amer, A.~Kulesza, and E.~Laenen,
  ``{NNLL resummation for stop pair-production at the LHC},''
  \href{http://dx.doi.org/10.1007/JHEP05(2016)153}{{\em JHEP} {\bfseries 05}
  (2016) 153}, \href{http://arxiv.org/abs/1601.02954}{{\ttfamily
  arXiv:1601.02954 [hep-ph]}}.

\bibitem{Borschensky:2020hot}
C.~Borschensky, B.~Fuks, A.~Kulesza, and D.~Schwartl\"ander, ``{Scalar
  leptoquark pair production at hadron colliders},''
  \href{http://dx.doi.org/10.1103/PhysRevD.101.115017}{{\em Phys. Rev. D}
  {\bfseries 101} no.~11, (2020) 115017},
  \href{http://arxiv.org/abs/2002.08971}{{\ttfamily arXiv:2002.08971
  [hep-ph]}}.

\bibitem{Alves:2002tj}
A.~Alves, O.~Eboli, and T.~Plehn, ``{Stop lepton associated production at
  hadron colliders},''
  \href{http://dx.doi.org/10.1016/S0370-2693(03)00266-1}{{\em Phys. Lett. B}
  {\bfseries 558} (2003) 165--172},
  \href{http://arxiv.org/abs/hep-ph/0211441}{{\ttfamily arXiv:hep-ph/0211441}}.

\bibitem{Dorsner:2014axa}
I.~Dorsner, S.~Fajfer, and A.~Greljo, ``{Cornering Scalar Leptoquarks at
  LHC},'' \href{http://dx.doi.org/10.1007/JHEP10(2014)154}{{\em JHEP}
  {\bfseries 10} (2014) 154}, \href{http://arxiv.org/abs/1406.4831}{{\ttfamily
  arXiv:1406.4831 [hep-ph]}}.

\bibitem{Hammett:2015sea}
J.~B. Hammett and D.~A. Ross, ``{NLO Leptoquark Production and Decay: The
  Narrow-Width Approximation and Beyond},''
  \href{http://dx.doi.org/10.1007/JHEP07(2015)148}{{\em JHEP} {\bfseries 07}
  (2015) 148}, \href{http://arxiv.org/abs/1501.06719}{{\ttfamily
  arXiv:1501.06719 [hep-ph]}}.

\bibitem{Mandal:2015vfa}
T.~Mandal, S.~Mitra, and S.~Seth, ``{Single Productions of Colored Particles at
  the LHC: An Example with Scalar Leptoquarks},''
  \href{http://dx.doi.org/10.1007/JHEP07(2015)028}{{\em JHEP} {\bfseries 07}
  (2015) 028}, \href{http://arxiv.org/abs/1503.04689}{{\ttfamily
  arXiv:1503.04689 [hep-ph]}}.

\bibitem{Schmaltz:2018nls}
M.~Schmaltz and Y.-M. Zhong, ``{The leptoquark Hunter\textquoteright{}s guide:
  large coupling},'' \href{http://dx.doi.org/10.1007/JHEP01(2019)132}{{\em
  JHEP} {\bfseries 01} (2019) 132},
  \href{http://arxiv.org/abs/1810.10017}{{\ttfamily arXiv:1810.10017
  [hep-ph]}}.

\bibitem{Faroughy:2016osc}
D.~A. Faroughy, A.~Greljo, and J.~F. Kamenik, ``{Confronting lepton flavor
  universality violation in B decays with high-$p_T$ tau lepton searches at
  LHC},'' \href{http://dx.doi.org/10.1016/j.physletb.2016.11.011}{{\em Phys.
  Lett. B} {\bfseries 764} (2017) 126--134},
  \href{http://arxiv.org/abs/1609.07138}{{\ttfamily arXiv:1609.07138
  [hep-ph]}}.

\bibitem{Raj:2016aky}
N.~Raj, ``{Anticipating nonresonant new physics in dilepton angular spectra at
  the LHC},'' \href{http://dx.doi.org/10.1103/PhysRevD.95.015011}{{\em Phys.
  Rev. D} {\bfseries 95} no.~1, (2017) 015011},
  \href{http://arxiv.org/abs/1610.03795}{{\ttfamily arXiv:1610.03795
  [hep-ph]}}.

\bibitem{Greljo:2017vvb}
A.~Greljo and D.~Marzocca, ``{High-$p_T$ dilepton tails and flavor physics},''
  \href{http://dx.doi.org/10.1140/epjc/s10052-017-5119-8}{{\em Eur. Phys. J. C}
  {\bfseries 77} no.~8, (2017) 548},
  \href{http://arxiv.org/abs/1704.09015}{{\ttfamily arXiv:1704.09015
  [hep-ph]}}.

\bibitem{Bansal:2018eha}
S.~Bansal, R.~M. Capdevilla, A.~Delgado, C.~Kolda, A.~Martin, and N.~Raj,
  ``{Hunting leptoquarks in monolepton searches},''
  \href{http://dx.doi.org/10.1103/PhysRevD.98.015037}{{\em Phys. Rev. D}
  {\bfseries 98} no.~1, (2018) 015037},
  \href{http://arxiv.org/abs/1806.02370}{{\ttfamily arXiv:1806.02370
  [hep-ph]}}.

\bibitem{Fuentes-Martin:2020lea}
J.~Fuentes-Martin, A.~Greljo, J.~Martin~Camalich, and J.~D. Ruiz-Alvarez,
  ``{Charm physics confronts high-p$_{T}$ lepton tails},''
  \href{http://dx.doi.org/10.1007/JHEP11(2020)080}{{\em JHEP} {\bfseries 11}
  (2020) 080}, \href{http://arxiv.org/abs/2003.12421}{{\ttfamily
  arXiv:2003.12421 [hep-ph]}}.

\bibitem{Allwicher:2022gkm}
L.~Allwicher, D.~A. Faroughy, F.~Jaffredo, O.~Sumensari, and F.~Wilsch,
  ``{Drell-Yan Tails Beyond the Standard Model},''
  \href{http://arxiv.org/abs/2207.10714}{{\ttfamily arXiv:2207.10714
  [hep-ph]}}.

\bibitem{Ohnemus:1994xf}
J.~Ohnemus, S.~Rudaz, T.~F. Walsh, and P.~M. Zerwas, ``{Single leptoquark
  production at hadron colliders},''
  \href{http://dx.doi.org/10.1016/0370-2693(94)90612-2}{{\em Phys. Lett. B}
  {\bfseries 334} (1994) 203--207},
  \href{http://arxiv.org/abs/hep-ph/9406235}{{\ttfamily arXiv:hep-ph/9406235}}.

\bibitem{Eboli:1997fb}
O.~J.~P. Eboli, R.~Zukanovich~Funchal, and T.~L. Lungov, ``{Signal and
  backgrounds for leptoquarks at the CERN LHC},''
  \href{http://dx.doi.org/10.1103/PhysRevD.57.1715}{{\em Phys. Rev. D}
  {\bfseries 57} (1998) 1715--1729},
  \href{http://arxiv.org/abs/hep-ph/9709319}{{\ttfamily arXiv:hep-ph/9709319}}.

\bibitem{Buonocore:2020erb}
L.~Buonocore, U.~Haisch, P.~Nason, F.~Tramontano, and G.~Zanderighi,
  ``{Lepton-Quark Collisions at the Large Hadron Collider},''
  \href{http://dx.doi.org/10.1103/PhysRevLett.125.231804}{{\em Phys. Rev.
  Lett.} {\bfseries 125} no.~23, (2020) 231804},
  \href{http://arxiv.org/abs/2005.06475}{{\ttfamily arXiv:2005.06475
  [hep-ph]}}.

\bibitem{Greljo:2020tgv}
A.~Greljo and N.~Selimovic, ``{Lepton-Quark Fusion at Hadron Colliders,
  precisely},'' \href{http://dx.doi.org/10.1007/JHEP03(2021)279}{{\em JHEP}
  {\bfseries 03} (2021) 279}, \href{http://arxiv.org/abs/2012.02092}{{\ttfamily
  arXiv:2012.02092 [hep-ph]}}.

\bibitem{Buonocore:2022msy}
L.~Buonocore, A.~Greljo, P.~Krack, P.~Nason, N.~Selimovic, F.~Tramontano, and
  G.~Zanderighi, ``{Resonant leptoquark at NLO with POWHEG},''
  \href{http://arxiv.org/abs/2209.02599}{{\ttfamily arXiv:2209.02599
  [hep-ph]}}.

\bibitem{Dorsner:2021chv}
I.~Doršner, S.~Fajfer, and A.~Lejlić, ``{Novel Leptoquark Pair Production at
  LHC},''
\href{http://arxiv.org/abs/2103.11702}{{\ttfamily arXiv:2103.11702 [hep-ph]}}.

\bibitem{Borschensky:2022xsa}
C.~Borschensky, B.~Fuks, A.~Jueid, and A.~Kulesza, ``{Scalar leptoquarks at the
  LHC and flavour anomalies: a comparison of pair-production modes at
  NLO-QCD},'' \href{http://arxiv.org/abs/2207.02879}{{\ttfamily
  arXiv:2207.02879 [hep-ph]}}.

\bibitem{Dorsner:2016wpm}
I.~Dor\v{s}ner, S.~Fajfer, A.~Greljo, J.~F. Kamenik, and N.~Ko\v{s}nik,
  ``{Physics of leptoquarks in precision experiments and at particle
  colliders},'' \href{http://dx.doi.org/10.1016/j.physrep.2016.06.001}{{\em
  Phys. Rept.} {\bfseries 641} (2016) 1--68},
  \href{http://arxiv.org/abs/1603.04993}{{\ttfamily arXiv:1603.04993
  [hep-ph]}}.

\bibitem{Alloul:2013bka}
A.~Alloul, N.~D. Christensen, C.~Degrande, C.~Duhr, and B.~Fuks, ``{FeynRules
  2.0 - A complete toolbox for tree-level phenomenology},''
  \href{http://dx.doi.org/10.1016/j.cpc.2014.04.012}{{\em Comput. Phys.
  Commun.} {\bfseries 185} (2014) 2250--2300},
  \href{http://arxiv.org/abs/1310.1921}{{\ttfamily arXiv:1310.1921 [hep-ph]}}.

\bibitem{Alwall:2014hca}
J.~Alwall, R.~Frederix, S.~Frixione, V.~Hirschi, F.~Maltoni, O.~Mattelaer,
  H.~S. Shao, T.~Stelzer, P.~Torrielli, and M.~Zaro, ``{The automated
  computation of tree-level and next-to-leading order differential cross
  sections, and their matching to parton shower simulations},''
  \href{http://dx.doi.org/10.1007/JHEP07(2014)079}{{\em JHEP} {\bfseries 07}
  (2014) 079}, \href{http://arxiv.org/abs/1405.0301}{{\ttfamily arXiv:1405.0301
  [hep-ph]}}.

\bibitem{NNPDF:2014otw}
{\bfseries NNPDF} Collaboration, R.~D. Ball {\em et~al.}, ``{Parton
  distributions for the LHC Run II},''
  \href{http://dx.doi.org/10.1007/JHEP04(2015)040}{{\em JHEP} {\bfseries 04}
  (2015) 040}, \href{http://arxiv.org/abs/1410.8849}{{\ttfamily arXiv:1410.8849
  [hep-ph]}}.

\bibitem{Borschensky:2021hbo}
C.~Borschensky, B.~Fuks, A.~Kulesza, and D.~Schwartl\"ander, ``{Scalar
  leptoquark pair production at the LHC: precision predictions in the era of
  flavour anomalies},'' \href{http://arxiv.org/abs/2108.11404}{{\ttfamily
  arXiv:2108.11404 [hep-ph]}}.

\bibitem{Langacker:1990jf}
P.~Langacker, ``{Parity violation in muonic atoms and cesium},''
  \href{http://dx.doi.org/10.1016/0370-2693(91)90688-M}{{\em Phys. Lett. B}
  {\bfseries 256} (1991) 277--283}.

\bibitem{Qweak:2018tjf}
{\bfseries Qweak} Collaboration, D.~Androi\'c {\em et~al.}, ``{Precision
  measurement of the weak charge of the proton},''
  \href{http://dx.doi.org/10.1038/s41586-018-0096-0}{{\em Nature} {\bfseries
  557} no.~7704, (2018) 207--211},
  \href{http://arxiv.org/abs/1905.08283}{{\ttfamily arXiv:1905.08283
  [nucl-ex]}}.

\bibitem{Barger:2000gv}
V.~D. Barger and K.-m. Cheung, ``{Atomic parity violation, leptoquarks, and
  contact interactions},''
  \href{http://dx.doi.org/10.1016/S0370-2693(00)00401-9}{{\em Phys. Lett. B}
  {\bfseries 480} (2000) 149--154},
  \href{http://arxiv.org/abs/hep-ph/0002259}{{\ttfamily arXiv:hep-ph/0002259}}.

\bibitem{Crivellin:2021egp}
A.~Crivellin, D.~M\"uller, and L.~Schnell, ``{Combined constraints on first
  generation leptoquarks},''
  \href{http://dx.doi.org/10.1103/PhysRevD.103.115023}{{\em Phys. Rev. D}
  {\bfseries 103} no.~11, (2021) 115023},
  \href{http://arxiv.org/abs/2104.06417}{{\ttfamily arXiv:2104.06417
  [hep-ph]}}. [Addendum: Phys.Rev.D 104, 055020 (2021)].

\bibitem{ParticleDataGroup:2020ssz}
{\bfseries Particle Data Group} Collaboration, P.~A. Zyla {\em et~al.},
  ``{Review of Particle Physics},''
  \href{http://dx.doi.org/10.1093/ptep/ptaa104}{{\em PTEP} {\bfseries 2020}
  no.~8, (2020) 083C01}.

\end{thebibliography}\endgroup
\end{document}